\documentclass[12pt]{article}
\usepackage[left=2.2cm, right=2.2cm, top=2.2cm, bottom=2.2cm]{geometry}
\usepackage{amsfonts}
\usepackage{amssymb}
\usepackage{amsmath}
\usepackage{lscape}
\usepackage{algorithm}
\usepackage[noend]{algpseudocode}
\usepackage{graphicx}

\newtheorem{theorem}{Theorem}

\newtheorem{corollary}{Corollary}

\newenvironment{proof}[1][Proof]{\noindent \textbf{#1.} }{\  \rule{0.5em}{0.5em}}

\newtheorem{example}{Example}

\makeatletter
\def\BState{\State\hskip-\ALG@thistlm}
\makeatother
\input{tcilatex}
\begin{document}

\title{On Consistency of Approximate Bayesian Computation\thanks{%
This research has been supported by Australian Research Council Discovery
Grant No. DP15010172. The authors are grateful to Judith Rousseau for very
helpful comments on an earlier draft of the paper.}}
\author{David T. Frazier\thanks{%
Department of Econometrics and Business Statistics, Monash University,
Melbourne, Australia. }, Gael M. Martin\thanks{%
Department of Econometrics and Business Statistics, Monash University,
Melbourne, Australia.} and Christian P. Robert\thanks{%
University of Paris Dauphine, Centre de Recherche en \'{E}conomie et
Statistique, and University of Warwick. Corresponding author; email:
xian@ceremade.dauphine.fr.}}
\maketitle

\begin{abstract}
Approximate Bayesian computation (ABC) methods have become increasingly
prevalent of late, facilitating as they do the analysis of intractable, or
challenging, statistical problems. With the initial focus being primarily on
the practical import of ABC, exploration of its formal statistical
properties has begun to attract more attention. The aim of this paper is to
establish general conditions under which ABC methods are Bayesian
consistent, in the sense of producing draws that yield a degenerate
posterior distribution at the true parameter (vector) asymptotically (in the
sample size). We derive conditions under which arbitrary summary statistics
yield consistent inference in the Bayesian sense, with these conditions\
linked to identification of the true parameters. Using simple illustrative
examples that have featured in the literature, we demonstrate that
identification, and hence consistency, is unlikely to be achieved in many
cases, and propose a simple diagnostic procedure that can indicate the
presence of this problem. We also formally explore the link between
consistency and the use of auxiliary models within ABC, and illustrate the
subsequent results in the\textbf{\ }Lotka-Volterra predator-prey
model.\bigskip \newline

\bigskip

\bigskip

\noindent \emph{MSC2010 Subject Classification}: 62F15, 62F12, 62C10
\end{abstract}

\newpage

\section{Introduction}

\baselineskip18pt

The use of approximate Bayesian computation (ABC) methods in models with
intractable likelihoods has gained increased momentum over recent years,
extending beyond the original applications in the biological sciences. (See
Marin \textit{et al., }2011, and Sisson and Fan, 2011, for recent reviews.).
Whilst ABC evolved initially as a practical tool, attention has begun to
shift to the investigation of its formal statistical properties, in
particular as they relate to the choice of summary statistics on which the
technique typically relies; see for example, Fearnhead and Prangle (2012),
Gleim and Pigorsch (2013), Marin \textit{et al. }(2014), Martin \textit{et
al. }(2014) and Martin \textit{et al. }(2014).

The aim of this paper is to establish general conditions under which summary
statistic-based ABC methods are Bayesian consistent, in the sense of
producing draws that yield a degenerate distribution at the true parameter
(vector) in the (sample size) limit. This aim is much broader than that
underlying Martin \textit{et al.} (2014),\textit{\ }in which standard
quasi-likelihood conditions were invoked to establish the Bayesian
consistency of auxiliary model-based versions of ABC. In particular, we
derive the conditions under which \textit{arbitrary }summary statistics
yield consistent inference, with these conditions linked to the
identification of the true parameters in any particular instance.\emph{\ }%
Using simple illustrative examples that have featured in the literature, we
demonstrate that consistency is \textit{not }achieved in many cases. This
finding calls into doubt routine applications of the ABC method that are
driven primarily by the convenience with which simple summary statistics can
be computed, without further thought being given to the information content
of those summaries.

Consistency by its very nature is more of a \textquotedblleft thought
experiment\textquotedblright\ than a practical feature of an estimation
procedure. Nonetheless, consistency is a useful metric with which to gauge
the output of a given statistical procedure. Following Diaconis and Freedman
(1986), we argue that regardless of Bayesian bearing, that is, whether one
is a \textquotedblleft Classical\textquotedblright\ Bayesian (who believes
in a \textquotedblleft true but unknown parameter which is to be estimated
from the data\textquotedblright ) or a \textquotedblleft
Subjective\textquotedblright\ Bayesian (who does not believe in true models
but, rather, thinks in terms of predictive distributions) consistency is
important for verification and practical implementation of Bayesian
procedures. That is, whilst consistency is a property that sits naturally
within the Classical\ Bayesian paradigm, it can also be viewed as being
important to Subjectivists. To wit, Blackwell and Dubins (1962) and Diaconis
and Freedman (1986) argue that consistency can be viewed as a
\textquotedblleft merging of intersubjective opinions\textquotedblright\ and
that consistency of the posterior\ implies that two separate subjective
Bayesians with different prior beliefs will ultimately end up with similar
predictive distributions.

In what follows, we only concern ourselves with the idea of consistency as
it pertains to some true model that is known up to an unknown vector of
parameters. In this setting Bayesian consistency means that any Bayesian
method should yield increasingly accurate\textbf{\ }posterior inference as
the sample size increases. While the theory of Bayesian consistency for
likelihood-based Bayesian methods is now well documented, at least in the
finite-dimensional parameter case, a thorough study on Bayesian consistency
of so-called likelihood free methods, such as ABC, has yet to be undertaken.
This represents an important gap in the literature, and is one we look to
fill.

Bayesian consistency for posteriors based on finite-dimensional parameters
is often derived under boundedness conditions for the underlying density
function of the true model; see, for example, Le Cam (1953), Ibragimov and
Has'minskii (1981), and Ghosal et al. (1995). In the ABC setting however,
conditions based on the underlying density function are not useful since by
the nature of the very problems to which ABC is applied, the underlying
density is typically unknown in closed form. To this end, we derive a set of
conditions on the summary statistics chosen within the ABC procedure that,
when satisfied, ensure consistency of the posterior obtained from ABC. These
conditions are similar in spirit to those seen in the literature on indirect
inference (Gouri\'{e}roux,\textit{\ et al.}, 1993). Examples from the ABC
literature are used to demonstrate how the aforementioned conditions can be
verified in practice.

The paper proceeds as follows. In Section \ref{abc} we briefly outline the
basic principles of ABC. In Section \ref{Aux}, we establish conditions under
which ABC will be consistent for the unknown parameters, and simple examples
that respectively do and do not satisfy these conditions are given. In
Section \ref{diag} we then propose a practical technique for identifying, in
any particular problem, when the conditions for consistency are (or are not)
satisfied. The analysis in Sections \ref{Aux} and \ref{diag} focuses on the
typical application of ABC, whereby summary statistics are chosen that are
deemed to contain some information about the parameters of the true model
and, more often than not, are used to define a matching criterion that is a
weighted function of sample moments. In Section \ref{crit} we couch the
discussion in terms of a general criterion function, where the latter
derives from an auxiliary model, and which may - but certainly does not need
to\ - derive from the likelihood function of that approximating model. In
Section \ref{ode_section} we pursue the matter of consistency when using ABC
to conduct inference in systems of ordinary differential equations (ODEs),
with the Lotka-Volterra system for predator and prey used for illustration,
and demonstrate that a common method for obtaining ABC posterior estimates
in this setting does not yield Bayesian consistent inference. Section \ref%
{disc} concludes. {Proofs of two theorems} and one corollary are provided in
an appendix to the paper.

\section{ABC: an Outline of the Basic Approach\label{abc}}

Suppose we are interested in conducting Bayesian inference on a complex
parametric model indexed by the unknown $p$-dimensional parameter $\mathbf{%
\theta }\in \mathbf{\Theta }$,\textbf{\ }$\mathbf{\Theta }\subset 
\mathbb{R}
^{p}$ compact, and let $P_{\mathbf{\theta }}$ denote the family of
probability measures induced by the model. Assume $P_{\mathbf{\theta }}$
admits a corresponding conditional density $p(\cdot |\mathbf{\theta })$ and
assume we have $T$ observations on the stochastic process $y_{t}$,
characterized by $p(\mathbf{y|\theta })$, with $\mathbf{y}%
=(y_{1},y_{2},...,y_{T})^{\prime }$\ denoting the $T$-dimensional vector of
observed data. The aim of ABC is to produce draws from an approximation to
the posterior distribution of the unknown $\mathbf{\theta }$ given observed
data $\mathbf{y}$, 
\begin{equation*}
p(\mathbf{\theta |y})\propto p(\mathbf{y|\theta })p(\mathbf{\theta }),
\end{equation*}%
in the case where both the prior, $p(\mathbf{\theta })$, and the likelihood, 
$p(\mathbf{y|\theta })$, can be easily simulated. These draws are used, in
turn, to approximate posterior quantities of interest, including marginal
posterior moments, marginal posterior distributions and predictive
distributions. The simplest (accept/reject) form of the algorithm (Tavar\'{e}%
\textit{\ et al., }1997, Pritchard \textit{et al.}, 1999) is detailed in
Algorithm \ref{ABC}. 
\begin{algorithm}
\caption{ABC algorithm}\label{ABC}
\begin{algorithmic}[1]
\State Simulate $\mathbf{\theta }^{i}$, $i=1,2,...,N$, from $p(\mathbf{\theta })$
\State Simulate $\mathbf{z}^{i}=(z_{1}^{i},z_{2}^{i},...,z_{T}^{i})^{\prime }$, $i=1,2,...,N$, from the likelihood, $p(\mathbf{.|\theta }^{i})$
\State Select $\mathbf{\theta }^{i}$ such that:%
\begin{equation}
d\{\mathbf{\eta }(\mathbf{y}),\mathbf{\eta }(\mathbf{z}^{i})\}\leq
\varepsilon ,  \label{distance}
\end{equation}%
where $\mathbf{\eta }(\mathbf{\cdot })$ is a (vector) statistic, $d\{\cdot
,\cdot \}$ is a distance function (or metric), and the tolerance level $%
\varepsilon $ is chosen as small as the computing budget allows.
\end{algorithmic}
\end{algorithm}

Algorithm \ref{ABC} thus samples $\mathbf{\theta }$ and $\mathbf{z}$ from
the joint\textit{\ }posterior:%
\begin{equation*}
p_{\varepsilon }(\mathbf{\theta },\mathbf{z|\eta (y)})=\frac{p(\mathbf{%
\theta })p(\mathbf{z|\theta })\mathbb{I}_{\varepsilon }[\mathbf{z}(\mathbf{%
\theta })]}{\textstyle\int_{\mathbf{\Theta }}\int_{\mathbf{z}}p(\mathbf{%
\theta })p(\mathbf{z|\theta })\mathbb{I}_{\varepsilon }[\mathbf{z}(\mathbf{%
\theta })]d\mathbf{z}d\mathbf{\theta }},
\end{equation*}%
where $\mathbb{I}_{\varepsilon }[\mathbf{z}(\mathbf{\theta })]$:=$\mathbb{I}%
[d\{\mathbf{\eta }(\mathbf{y}),\mathbf{\eta }(\mathbf{z(\theta )})\}\leq
\varepsilon ]$ is one if $d\left\{ \mathbf{\eta }(\mathbf{y}),\mathbf{\eta }(%
\mathbf{z(\theta )})\right\} \leq \varepsilon $ and zero else.\footnote{%
The notation $\mathbf{z(\theta )}$\ is used to emphasize the dependence of
the simulated $\mathbf{z}$\ on $\mathbf{\theta }.$} Clearly, when $\mathbf{%
\eta }(\mathbf{y })$ is a sufficient statistic and $\varepsilon $ is
arbitrarily small,%
\begin{equation}
p_{\varepsilon }(\mathbf{\theta |\eta (y)})=\textstyle\int_{\mathbf{z}%
}p_{\varepsilon }(\mathbf{\theta },\mathbf{z|\eta (y)})d\mathbf{z}
\label{abc_post}
\end{equation}%
approximates the\textbf{\ }exact posterior, $p(\mathbf{\theta |y})$, and
draws from $p_{\varepsilon }(\mathbf{\theta },\mathbf{z|\eta (y)})$ can be
used to estimate features of the true posterior. In practice however, the
complexity of the models to which ABC is applied implies, almost by
definition, that sufficiency is unattainable. Hence, in the limit, as $%
\varepsilon \rightarrow 0$, the draws can be used only to approximate
features of $p(\mathbf{\theta |\eta }(\mathbf{y})).$

ABC-based estimates of $p(\mathbf{\theta |y})$ thus suffer from three types
of approximation error: one invoked by the use of summary statistics that
are not sufficient for\textbf{\ }$\mathbf{\theta }$;\textbf{\ }another
associated with the use in practice of a non-zero tolerance, $\varepsilon $,
for selecting draws from\textbf{\ }$p(\mathbf{\theta |\eta }(\mathbf{y}))$;%
\textbf{\ }and, thirdly, the error produced when using non-parametric
density techniques to estimate $p(\mathbf{\theta |\eta (y)})$\ from a given
set of selected draws. For any level of overall computational burden (i.e.,
the total number of draws $N$), reducing $\varepsilon $ comes at a cost of
reducing the probability of a draw being accepted, thereby contributing to
the third form of error. The problem is exacerbated the larger is the
dimension of\textbf{\ }$\mathbf{\eta }(\mathbf{y})$; see Blum (2010), Blum 
\textit{et al. }(2013) and Nott \textit{et al. }(2014). In practice $%
\varepsilon $\ tends to be chosen such that, for a given value of $N$, a
certain (small) proportion of draws of $\mathbf{\theta }^{i}$\ are selected,
with attempts then made to reduce the third form of error using a variety of
post-sampling (kernel-based) corrections of the draws (Beaumont \textit{et
al.}, 2002, Blum, 2010, Blum and Fran\c{c}ois, 2010). Other work gives
emphasis to choosing $\mathbf{\eta (\cdot )}$ and/or the selection mechanism
itself in such a way that $p(\mathbf{\theta |\eta }(\mathbf{y}))$ is\textbf{%
\ }a closer match to $p(\mathbf{\theta |y})$, in some sense. This may
involve the replacement of the basic accept/reject scheme with Markov chain
Monte Carlo (MCMC) and/or sequential Monte Carlo (SMC) steps (Marjoram 
\textit{et al.}, 2003, Sisson \textit{et al.,} 2007, Beaumont \textit{\ et
al.}, 2009, Toni \textit{et al.}, 2009 and Wegmann \textit{et al.}, 2009);
or the selection of a vector $\mathbf{\eta (\cdot )}$ that is more
informative in some well-defined sense; see Joyce and Marjoram (2008),
Wegmann \textit{et al.} (2009), Blum (2010) and Fearnhead and Prangle (2012).

In this latter spirit - and mimicking the frequentist techniques of indirect
inference (II) (Gouri\'{e}roux \textit{et al.}, 1993, Heggland and Frigessi,
2004) and efficient method of moments (EMM) (Gallant and Tauchen, 1996),
Drovandi \textit{et al.} (2011), Gleim and Pigorsch (2013), Martin \textit{%
et al. }(2014), Drovandi \textit{et al.} (2015) and Creel and Kristensen
(2015) exploit an approximating model to produce the summary statistic
vector $\mathbf{\eta }(\mathbf{\cdot })$. Under certain conditions on the
auxiliary model, asymptotic sufficiency (at least) is attainable via use of
the maximum likelihood estimates of the auxiliary parameters as the matching
statistics in the ABC algorithm.\textbf{\ }Martin \textit{et al.} also prove
(for $\varepsilon \rightarrow 0$) the (Bayesian) consistency of the ABC
approach that\textbf{\ }uses the MLE of the parameters of the auxiliary
model to define\textbf{\ }$\mathbf{\eta }(\mathbf{\cdot })$, under similar
conditions to those used to prove the consistency of the II method. The
authors demonstrate the equivalence (again, as the tolerance approaches
zero) of inference based on the score of the auxiliary model to that based
on the MLE. This equivalence holds for any sample size and, hence, ensures
that consistency is maintained by the (computationally efficient)
score-based approach on the satisfaction of the appropriate conditions.

In this paper we also address the issue of Bayesian consistency, but in the
completely general setting in which\textbf{\ }$\mathbf{\eta (\cdot )}$\
comprises an arbitrary vector summary statistic, with elements possibly
including, but not limited to, sample moments of the data, and\textbf{\ }%
with $\mathbf{\eta (\cdot )}$ not necessarily having an explicit link to the
parameters of an auxiliary model.\textbf{\ }In the particular situation where%
\textbf{\ }$\mathbf{\eta (}\mathbf{\cdot )}$\textbf{\ }forms a vector
statistic composed of sample moments, ABC parallels the frequentist method
of simulated moments (McFadden, 1989, Pakes and Pollard, 1989, Duffie and
Singleton, 1993). In the following section we maintain full generality in
terms of the definition of $\mathbf{\eta (\cdot ).}$ In Section \ref{crit}
we then consider the case where the matching criterion is explicitly defined
with respect to an auxiliary model, highlighting the fact that the
likelihood function of that model is by no means the only possible criterion
that can be adopted.

\section{ABC and Consistency\label{Aux}}

\subsection{Consistency and Summary Statistics\label{Theory}}

Herein, we will only concern ourselves with the Classical\ ideal of Bayesian
consistency: namely, as more data accumulates the posterior should stabilize
around some true value and eventually collapse to a point mass at the same
true value. More formally, for some set $A\subset \Theta $, define the
posterior probability of $A$ as\textbf{\ \ }%
\begin{equation*}
\text{Pr}(\mathbf{\theta }\in A|\mathbf{y})=\displaystyle\int\limits_{A}p(%
\mathbf{\theta |y})d\mathbf{\theta ,}
\end{equation*}%
we then have the following {well-known} definition:

\begin{description}
\item[Definition 1:] For true value\textbf{\ }$\mathbf{\theta }=\mathbf{%
\theta }^{0},$ the posterior density $p(\mathbf{\theta |y})$ is \textit{%
Bayesian consistent}\textbf{\ }if for any $\delta >0$\textbf{\ }and $\aleph
_{\delta }(\mathbf{\theta }^{0})$\textbf{\ }an open neighborhood of\textbf{\ 
}$\mathbf{\theta }^{0}$,\textbf{\ }$\text{Pr}\mathbf{(\theta \notin \aleph }%
_{\delta }\mathbf{(\theta }^{0}\mathbf{)|y)}\overset{P}{\rightarrow }0$%
\textbf{\ }as\textbf{\ }$T\rightarrow \infty .$
\end{description}

\noindent%
Herein, the symbol $\overset{P}{\rightarrow }$ denotes convergence in
probability, and the symbols $o_{P}(a_{T}),O_{P}(b_{T}),\text{plim}%
_{T\rightarrow \infty },$ to be used below, have the usual definition.

Unlike the notion of consistency defined above, Bayesian consistency of
posterior densities obtained from ABC requires not only\textbf{\ }$%
T\rightarrow \infty $ but\textbf{\ }$\varepsilon \rightarrow 0$ and is
particular to the choice of $\mathbf{\eta }(\cdot )$ \textbf{(}and, indeed%
\textbf{\ }$d\{\mathbf{\cdot ,\cdot }\}$).\textbf{\ }Given this fact, we
require a separate definition of Bayesian consistency for ABC.

\begin{description}
\item[Definition 2:] For true value\textbf{\ }$\mathbf{\theta }=\mathbf{%
\theta }^{0}$ and (vector) summary statistics\textbf{\ }$\mathbf{\eta (\cdot
);}$ $%
\mathbb{R}
^{T}\rightarrow \mathbf{B}$,\textbf{\ }where\textbf{\ }$\mathbf{B}\subset 
\mathbf{%
\mathbb{R}
}^{d}\ $and$\ d\geq p,$ {the ABC-based posterior density} $p_{\varepsilon }(%
\mathbf{\theta |\eta (y)})$\ is \textit{Bayesian consistent} if for any $%
\delta >0$,\textbf{\ }$\text{Pr}_{\varepsilon }(\mathbf{\theta \notin \aleph
_{\delta }(\theta ^{0})|\eta (y)})\overset{P}{\rightarrow }0$\textbf{\ }as%
\textbf{\ }$T\rightarrow \infty $ and $\varepsilon \rightarrow 0$.
\end{description}

\noindent In addition, and in common with the standard definition (Defn. 1),
the prior density used in ABC must be positive at the true value $\mathbf{%
\theta }^{0}$ and so we will assume the following condition is satisfied.

\begin{description}
\item[{Assumption [P]:}] The prior density $p(\cdot )$ is continuous and $p(%
\mathbf{\theta }^{0})>0.$
\end{description}

\medskip

\begin{figure}[h!]
	\centering
\includegraphics[width=4.4754in,height=2.5936in]{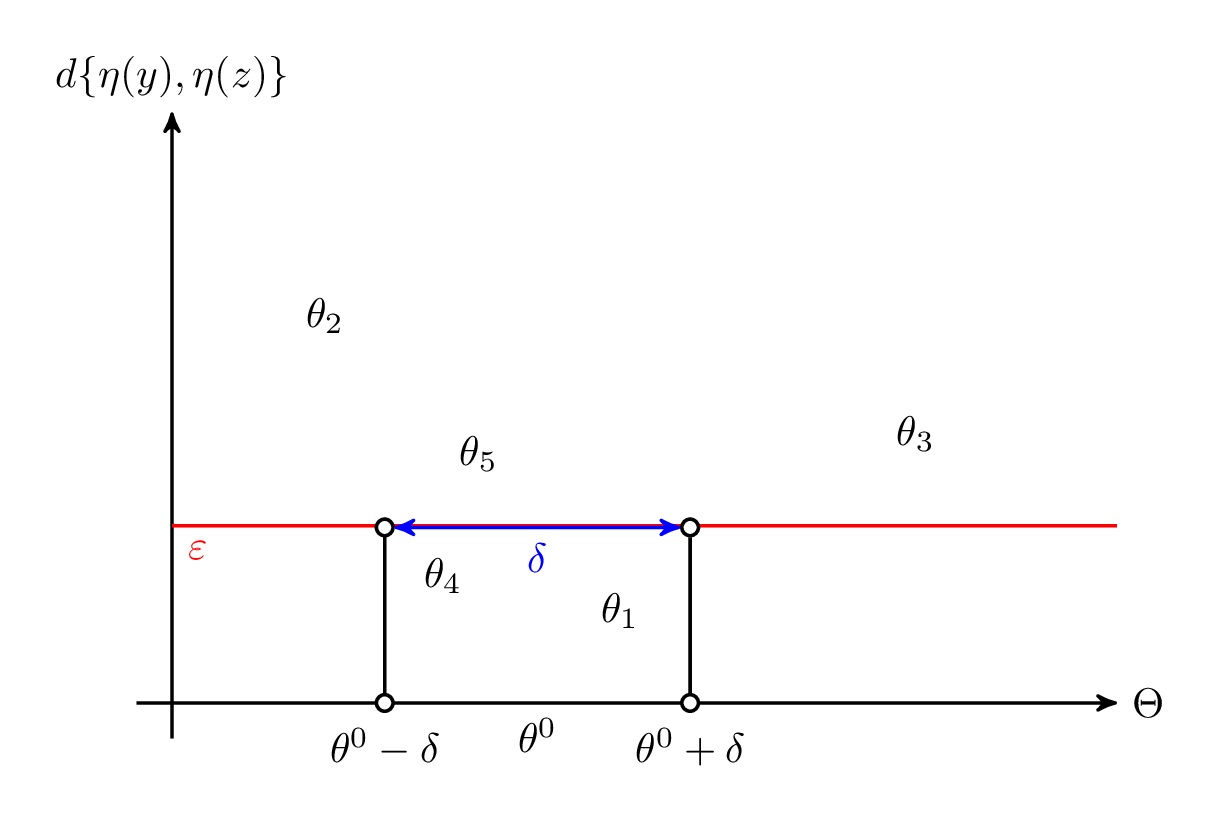}
\caption{An illustration
of\textbf{\ }ABC output for $N=5$ simulations and given values of $\theta ^{0},\protect\delta ,\protect\varepsilon ,$ and $\mathbf{\protect\eta }(\mathbf{y}).$}
\label{simplefig1}
\end{figure}

As a heuristic for what Bayesian consistency in the ABC setting entails,
consider the following simple example. Assume $\Theta :=[0,1]$ and that $%
\theta \in \Theta $ has uniform prior probability. Consider some $\delta >0$%
, $\varepsilon >0$, and assume we have data $\mathbf{y}$ that is generated
according to true (scalar) value $\theta ^{0}$. For $N=5$ simulations the
output is contained in Figure \ref{simplefig1}. For the particular $\delta
>0 $ and $\varepsilon >0$ chosen, two points lie within $\aleph _{\delta
}(\theta ^{0})$ and three points lie outside. Clearly, the ABC-based
posterior density $p_{\varepsilon }(\theta \mathbf{|\eta (y)})$ only places
mass on $\theta _{1}$ and $\theta _{4}$, as these points lead to a distance
less than $\varepsilon $, and zero mass is placed on the remaining three
points. The ABC-based posterior density $p_{\varepsilon }(\theta \mathbf{%
|\eta (y)})$ will be Bayesian consistent if for any arbitrary $\delta >0$
and some $\varepsilon >0$ similar behavior to that observed in Figure \ref%
{simplefig1} holds as $T\rightarrow \infty $. This requires the following to
be satisfied: one, for any given $\delta >0$, we must be able to simulate
draws within $\aleph _{\delta }(\theta ^{0})$ (guaranteed by Assumption
[P]); two, for any $T$, including\textbf{\ }large $T$, there must exist a
value of $\varepsilon $ such that the only draws satisfying $d\{\mathbf{\eta
(y),\eta (z)}\}\leq \varepsilon $ are those in $\aleph _{\delta }(\theta
^{0})$; three, for the value of $\varepsilon $ in two, there must exist a
corresponding number of simulation draws $N(\varepsilon )$ such that at
least one simulated $\theta ^{i}\in \{\theta ^{i}\}_{i=1}^{N(\varepsilon )}$
satisfying $d\{\mathbf{\eta (y),\eta }(\mathbf{z}(\theta ^{i}))\}\leq
\varepsilon $ occurs, else $p_{\varepsilon }(\theta \mathbf{|\eta (y)})$
will not exist. 
The formalization of these statements, along with the precise set of
assumptions that a vector of summary statistics, $\mathbf{\eta (y)},$\textbf{%
\ }should satisfy in order for ABC to yield consistent inference, is the
content of Theorem 1 and its proof. Subsequent to the presentation of the
theorem, we provide a simple example in which the conditions are satisfied,
followed by a second example in which they are not. The way in which an
increase in the dimension of $\mathbf{\eta (y)}$ can be used to retrieve
consistency in the latter case, is then illustrated.

Define the limiting value of the summary statistic based on observed data
(respectively, simulated data) as\textbf{\ }$\mathbf{b(\theta }^{0}\mathbf{)}
$\textbf{\ (}respectively,\textbf{\ }$\mathbf{b(\theta }^{i}\mathbf{)}$),
and let $\Vert \cdot \Vert $ denote the Euclidean norm.

\begin{theorem}
\label{thm1} Let $d\{\cdot ,\cdot \}$ be an induced metric on the normed
space $(\mathbf{B,}\left\Vert \cdot \right\Vert _{\ast })$. Given summary
statistics $\mathbf{\eta }(\mathbf{y})$, assume that\textbf{\ }the following
conditions\textbf{\ }are satisfied:

\begin{description}
\item[{[S0]}] The DGP for $\mathbf{y}$ is uniquely defined at $\mathbf{\theta 
}^{0}.$

\item[{[S1]}] $\Vert \mathbf{\eta }(\mathbf{y})-\mathbf{b}(\mathbf{\theta }%
^{0})\Vert =o_{P}(1)$.

\item[{[S2]}] The map $\mathbf{\theta }^{i}\mapsto \mathbf{b}(\mathbf{\theta }%
^{i})$ is deterministic, continuous, and satisfies

\begin{description}
\item[{[S2(1)]}] $\sup_{\mathbf{\theta \in \Theta }}\Vert \mathbf{\eta }(%
\mathbf{z}(\mathbf{\theta }))-\mathbf{b}(\mathbf{\theta })\Vert =o_{P}(1)$.

\item[{[S2(2)]}] $\mathbf{b}(\mathbf{\cdot })$ is one-to-one in $\mathbf{%
\theta }^{i}$.
\end{description}
\end{description}

\noindent%
If [S0]-[S2] above and [P] are satisfied, then, for any $\delta >0,$ $\text{%
Pr}_{\varepsilon }(\mathbf{\theta }\notin \aleph _{\delta }(\mathbf{\theta }%
^{0})|\mathbf{\eta (y))}\overset{P}{\rightarrow }0$ as $T\rightarrow \infty $
and $\varepsilon \rightarrow 0.$
\end{theorem}

\begin{description}
\item[Remark 1] Theorem \ref{thm1} states that, for ABC based on $\mathbf{%
\eta }(\cdot )$ to be consistent the limit map $\mathbf{\theta }^{i}\mapsto 
\mathbf{b}(\mathbf{\theta }^{i})$ must act in the same manner as the
\textquotedblleft binding function\textquotedblright\ in indirect inference;
see Gouri\'{e}roux \textit{et al.} (1993) and Gouri\'{e}roux and Monfort
(1996) for a general discussion of binding functions.

\item[Remark 2] Since we are only concerned with Classical\ Bayesian
consistency, Assumption [S0] is implicit and therefore not explicitly
required. However, Assumption [S0] is a deep identification condition that
may not be satisfied in all circumstances and is therefore maintained to
illustrate the scope of the models to which this result will apply.
Assumption [S1] is often satisfied under general conditions restricting the
dependence in the observed data. Assumption [S2(1)] requires that for all $%
\xi >0,$ 
\begin{equation*}
\lim_{T\rightarrow \infty }\Pr \left( \sup_{\mathbf{\theta \in \Theta }%
}\left\Vert \mathbf{\eta }(\mathbf{z(\theta )})-\mathbf{b}(\mathbf{\theta }%
)\right\Vert >\xi \right) =0,
\end{equation*}%
and is generally referred to as uniform convergence. This stronger notion of
convergence is required to ensure that the simulated paths $\mathbf{z(\theta
)}$, and the subsequent $\mathbf{\eta }(\mathbf{z(\theta )})$,\ are
well-behaved over $\mathbf{\Theta }$. \ General conditions determining
satisfaction of [S2(1)] are now well-known and a great many results can be
obtained from the empirical process literature; see, for instance, Pollard
(1990). In particular, [S2(1)] is likely to be satisfied for many different
types of summary statistics so long as the prior density $p(\cdot )$ admits
values of\textbf{\ }$\mathbf{\theta }$ that do not allow the simulated data
to display too much persistence.\footnote{%
Technically, conditions [S0] and [S2(1)] imply condition [S1]. However, the
authors believe it is helpful to specify separate conditions on the
statistics associated with observed and simulated data.}

\item[Remark 3] Theorem \ref{thm1} requires that the (vector of) summary
statistics based on observed data converges, with respect to\textbf{\ }$%
d\{\cdot ,\cdot \},$ to a fixed quantity and the corresponding vector of
statistics based on simulated data $\mathbf{z}^{i}=\mathbf{z(\theta }^{i}%
\mathbf{)}$ converges (uniformly), with respect to\textbf{\ }$d\{\cdot
,\cdot \},$ to a deterministic function of $\mathbf{\theta }^{i}$.
Consistency thus depends not only on the choice of\textbf{\ }$\mathbf{\eta
(y)}$\textbf{\ }but also on the precise choice of\textbf{\ }$d\{\cdot ,\cdot
\},$ with convergence in one metric not necessarily implying convergence in
another. However, restricting\textbf{\ }$d:\mathbf{B}\times \mathbf{B}%
\rightarrow 
\mathbb{R}
_{+}$ to be an induced metric$\ $on the normed space $(\mathbf{B,}\left\Vert
\cdot \right\Vert _{\ast })$ -\ i.e., for $\mathbf{\eta }_{1}\mathbf{,\eta }%
_{2}\in \mathbf{B\,}$, requiring that\textbf{\ }$d\{\mathbf{\eta }_{1}%
\mathbf{,\eta }_{2}\}=$ $\left\Vert \mathbf{\eta }_{1}\mathbf{-\eta }%
_{2}\right\Vert _{\ast }$ for some norm $\left\Vert \cdot \right\Vert _{\ast
}$ - relieves the convergence issue since all norms on\textbf{\ }$\mathbf{B}$
are equivalent to the Euclidean norm\textbf{\ $\Vert \cdot \Vert $}.\textbf{%
\ }The requirement that\textbf{\ }$d\{\cdot ,\cdot \}$ be an induced metric
is not restrictive as the most common choices of $d\{\cdot ,\cdot \}$\
satisfy this condition.

\item[Remark 4] Bayesian consistency says that for any $\delta >0$, $%
p_{\varepsilon }(\mathbf{\theta }|\mathbf{\eta (y)})$ will attribute zero
probability, as $T\rightarrow \infty $, to points outside $\aleph _{\delta }(%
\mathbf{\theta }^{0})$; it does not say \textit{anything} about how well $%
p_{\varepsilon }(\mathbf{\theta }|\mathbf{\eta (y)})$ approximates the
posterior density $p(\mathbf{\theta |y})$ or even the partial posterior
density $p(\mathbf{\theta }|\mathbf{\eta (y)})$. Specifically, the
demonstration of Bayesian consistency is distinct from existing theoretical
work on ABC that shows $p_{\varepsilon }(\mathbf{\theta }|\mathbf{\eta (y)})$
is consistent for $p(\mathbf{\theta }|\mathbf{\eta (y)})$, as $N\rightarrow
\infty $ and $\varepsilon \rightarrow 0,$ for any $\mathbf{\theta \in \Theta 
}$ and for any fixed T$\mathbf{.}$ To prove the latter form of result,
researchers have borrowed from the literature on nonparametric density
estimation and relied on the idea of mean squared error (MSE) consistency,
which requires the bias and variance of $p_{\varepsilon }(\mathbf{\theta }|%
\mathbf{\eta (y)})-$ $p(\mathbf{\theta }|\mathbf{\eta (y)})$ to approach zero%
\textbf{\ }as $N\rightarrow \infty $ and $\varepsilon \rightarrow 0;$ see,
for example, Blum (2010) and Biau \textit{et al.} (2015). In particular, MSE
consistency requires a specific rate condition between\textbf{\ }$N\ $and%
\textbf{\ }$\varepsilon $\textbf{\ }to ensure that the variance of $%
p_{\varepsilon }(\mathbf{\theta }|\mathbf{\eta (y)})-$ $p(\mathbf{\theta }|%
\mathbf{\eta (y)})$ shrinks to zero fast enough. As noted in the proof of
Theorem \ref{thm1}, Bayesian consistency still requires $N$ to increase as $%
\varepsilon \rightarrow 0,$ but only to ensure that\textbf{\ }$%
p_{\varepsilon }(\mathbf{\theta |\eta (y)})$ exists for small\textbf{\ }$%
\varepsilon $, and any $T$. In this way, the particular relationship between 
$N$ and $\varepsilon $ is independent of the sample size $T$. {This lack of
any }$T$-dependent condition for $\varepsilon $ {contrasts with the need for
such a condition when deriving results for the asymptotic distribution of
ABC point estimators; see, for example, Li and Fearnhead (2015). We
elaborate further on this distinction in an on-line supplementary appendix
to the paper.\footnote{%
This document is available at:{\ }http://users.monash.edu.au/\symbol{126}%
gmartin/FMR\_Supplementary\_Appendix.pdf).}}
\end{description}

\subsection{Success and Failure of Summary Statistic-based ABC}

Consistency of ABC based on $\mathbf{\eta }(\cdot )$ hinges on the
particular form of $\mathbf{b}(\mathbf{\theta }^{i})$. If $\mathbf{b}(%
\mathbf{\cdot })$ is one-to-one, i.e., the map $\mathbf{\theta }^{i}\mapsto 
\mathbf{b}(\mathbf{\theta }^{i})$ satisfies [S2(2)], and the remaining
assumptions in\textbf{\ }Theorem\textbf{\ }\ref{thm1}\textbf{\ }hold,\textbf{%
\ }ABC based on $\mathbf{\eta }(\cdot )$ will be consistent. There is
generally no guarantee that $\mathbf{b(\cdot )}$ will be one-to-one and
satisfaction of this condition depends on both\textbf{\ }the true structural
model and the particular choice of summary statistics. Examples 1 and 2
illustrate a case where [S2(2)] is and is not satisfied, respectively.
Example 3 illustrates the impact on identification and, hence, the
attainment of consistency, of adding summary statistics to an initial set.

\begin{example}[Satisfaction of S2(2) ]
Consider the following autoregressive (AR) model of order one: 
\begin{equation*}
y_{t}=\theta y_{t-1}+\nu _{t},
\end{equation*}%
where $\nu _{t}\sim $i.i.d.$N(0,1)$ and $|\theta |<1,\theta \neq 0$. Whilst
the likelihood for this model is known in closed form and, hence, exact
Bayesian inference is perfectly feasible, for the sake of illustration,
consider Algorithm \ref{ar1_ABC} based on the summary statistic $\eta (%
\mathbf{y})=\frac{1}{T}\sum_{t=2}^{T}y_{t}y_{t-1}$.

\begin{algorithm}
\caption{ABC algorithm: AR(1) example}\label{ar1_ABC}
\begin{algorithmic}[1]
\State Simulate the AR(1) coefficient $\theta ^{i}$ from, for
instance, a uniform prior over $(-1,1);$
\State Generate an i.i.d. sequence $\{\nu _{t}\}_{t=1}^{T}$;
\State Produce a simulated series $\{z_{t}^{i}(\theta
^{i})\}_{t=1}^{T}$;
\State Accept\textbf{\ }the value simulated in Step (1) if $d\{\eta
(\mathbf{y}),\eta (\mathbf{z}^{i})\}\leq \varepsilon $ for $\varepsilon >0$
and small.
\end{algorithmic}
\end{algorithm} Assume that some true value $\theta ^{0}$ has generated the
observed sample $\mathbf{y}$. For $p_{\varepsilon }\left( \theta |\eta (%
\mathbf{y})\right) $ to be degenerate at $\theta ^{0}$ it must be that $%
d\{b(\theta ^{0}),b(\theta ^{i})\}=0$ has a\textbf{\ }unique solution $%
\theta ^{i}=\theta ^{0}$. By the weak law of large numbers $\eta (\mathbf{y})%
\overset{P}{\rightarrow }E[y_{t}y_{t-1}]=$ $b(\theta ^{0})=\theta
^{0}/(1-(\theta ^{0})^{2})$ and\textbf{\ }$\eta (\mathbf{z}^{i})\overset{P}{%
\rightarrow }E[z_{t}^{i}z_{t-1}^{i}]=$ $b(\theta ^{i})=\theta
^{i}/(1-(\theta ^{i})^{2})$, so $d\{b(\theta ^{0}),b(\theta ^{i})\}=0$
requires that 
\begin{equation*}
0=b(\theta ^{0})-b(\theta ^{i})=(\theta ^{i})^{2}\theta ^{0}+\theta
^{i}(1-(\theta ^{0})^{2})-\theta ^{0}
\end{equation*}%
has unique solution $\theta ^{i}=\theta ^{0}$. This quadratic equation in $%
\theta ^{i}$ has two solutions: $\theta ^{i}=\theta ^{0}$ and $\theta
^{i}=-1/\theta ^{0}$. However, given that $|\theta ^{0}|<1,\;\theta ^{0}\neq
0$, the second solution is not in the feasible region for $\theta ^{i}$ and
so ABC based on $\eta (\mathbf{y})=\frac{1}{T}\sum_{t=2}^{T}y_{t}y_{t-1}$
satisfies the conditions of Theorem \ref{thm1}.
\end{example}

\begin{example}[Failure of S2(2)]
Consider now the moving average (MA) model of order two: 
\begin{equation}
y_{t}=e_{t}+\theta _{1}e_{t-1}+\theta _{2}e_{t-2},  \label{MA2}
\end{equation}%
where $e_{t}\sim i.i.d.N(0,1)$ and $\theta _{1},\theta _{2}$ satisfy the
following invertibility conditions 
\begin{equation}
-2<\theta _{1}<2,\;\theta _{1}+\theta _{2}>-1,\theta _{1}-\theta _{2}<1.
\label{const1}
\end{equation}%
Following Marin \textit{et al.} (2011), we choose as summary statistics the
sample autocovariances $\eta _{j}(\mathbf{y})=\frac{1}{T}%
\sum_{t=1+j}^{T}y_{t}y_{t-j}$, for $j=0,1,2...,K.$ Consider, initially,
Algorithm \ref{ma2_ABC}, based on $\mathbf{\eta }\left( \mathbf{y}\right)
=(\eta _{0}\mathbf{(y)}, \eta _{1}\mathbf{(y)})^{\prime }$.

\begin{algorithm}
\caption{ABC algorithm: MA(2) example}\label{ma2_ABC}
\begin{algorithmic}[1]
\State Simulate the MA(2) coefficients $\mathbf{\theta }^{i\text{ }%
} $from $p\left( \mathbf{\theta }\right) $ satisfying \eqref{const1}, where%
\textbf{\ }$\mathbf{\theta }=(\theta _{1},\theta _{2})^{\prime }$;
\State Generate an i.i.d. sequence $\{e_{t}\}_{t=1}^{T}$;
\State Produce a simulated series $\{z_{t}^{i}\left( \mathbf{%
\theta }^{i}\right) \}_{t=1}^{T}$;
\State Accept\textbf{\ }the value generated in Step (1) if $d\{\mathbf{%
\eta (y)},\mathbf{\eta }(\mathbf{z}^{i})\}\leq \varepsilon $ for $%
\varepsilon >0$ and small.
\end{algorithmic}
\end{algorithm} Assume that true value $\mathbf{\theta }^{0}=(\theta
_{1}^{0},\theta _{2}^{0})^{\prime }$ has generated the observed data $%
\mathbf{y}$. By the \textit{weak law }of large numbers $\eta _{0}(\mathbf{y})%
\overset{P}{\rightarrow }E[y_{t}^{2}]=1+(\theta _{1}^{0})^{2}+(\theta
_{2}^{0})^{2}$ and $\eta _{1}(\mathbf{y})\overset{P}{\rightarrow }%
E[y_{t}y_{t-1}]=\theta _{1}^{0}(1+\theta _{2}^{0})$.\textbf{\ }\textit{In
addition, }conditional on $\mathbf{\theta }^{i}=(\theta _{1}^{i},\theta
_{2}^{i})^{\prime }$ satisfying equation \eqref{const1}, $\eta _{0}(\mathbf{z%
}^{i})\overset{P}{\rightarrow }1+(\theta _{1}^{i})^{2}+(\theta _{2}^{i})^{2}$
and $\eta _{1}(\mathbf{z}^{i})\overset{P}{\rightarrow }\theta
_{1}^{i}(1+\theta _{2}^{i}).$ For $p_{\varepsilon }\left( \mathbf{\theta }|%
\mathbf{\eta (y)}\right) $ obtained from the above algorithm to be
degenerate at $\mathbf{\theta }^{0}$ it must be that for all $\mathbf{\theta 
}^{i}\mathbf{,\theta }^{0}\mathbf{\in \Theta }$, $0=\mathbf{b}(\mathbf{%
\theta }^{0})-\mathbf{b}(\mathbf{\theta }^{i})$ has unique solution $\mathbf{%
\theta }^{i}=\mathbf{\theta }^{0}.$ Clearly,%
\begin{equation*}
0=\mathbf{b}(\mathbf{\theta }^{0})-\mathbf{b}\left( \mathbf{\theta }%
^{i}\right) =%
\begin{pmatrix}
1+(\theta _{1}^{0})^{2}+(\theta _{2}^{0})^{2} \\ 
\theta _{1}^{0}(1+\theta _{2}^{0})%
\end{pmatrix}%
-%
\begin{pmatrix}
1+(\theta _{1}^{i})^{2}+(\theta _{2}^{i})^{2} \\ 
\theta _{1}^{i}(1+\theta _{2}^{i})%
\end{pmatrix}%
.
\end{equation*}%
\newline
As in Marin et al. (2011), take $\theta _{1}^{0}=.6,\theta _{2}^{0}=.2$.
Then the question becomes, does there exist $\theta _{1}^{i}\neq .6,\theta
_{2}^{i}\neq .2$ such that 
\begin{equation}
0=\mathbf{b}(\theta ^{0})-\mathbf{b}\left( \theta ^{i}\right) =%
\begin{pmatrix}
1+(.6)^{2}+(.2)^{2} \\ 
.6(1+.2)%
\end{pmatrix}%
-%
\begin{pmatrix}
1+(\theta _{1}^{i})^{2}+(\theta _{2}^{i})^{2} \\ 
\theta _{1}^{i}(1+\theta _{2}^{i})%
\end{pmatrix}%
?  \label{Truth}
\end{equation}%
Simple numerical calculations reveal that \eqref{Truth} has two solutions: $%
\theta _{1}^{i}=.6,\theta _{2}^{i}=.2$ and $\theta _{1}^{i}\approx
.5453,\theta _{2}^{i}\approx .3204$, where the latter solution remains in
the feasible region for $\mathbf{\theta }^{i}=(\theta _{1}^{i},\theta
_{2}^{i})^{\prime }.$ Therefore, $\mathbf{b}(\mathbf{\cdot })$ is not
one-to-one and the \textit{ABC-based p}osterior will not converge to $%
\mathbf{\theta }^{0}=(.6,.2)^{\prime }$.
\end{example}

\begin{example}[Effect of Additional Statistics]
Consider the same MA(2) model as in Example 2, but now consider the use of
the three-dimensional vector of summary statistics:\textbf{\ }%
\begin{equation*}
\mathbf{\eta }\left( \mathbf{y}\right) =(\eta _{0}\mathbf{(y)}, \eta _{1}%
\mathbf{(y)},\eta _{2}\mathbf{(y)})^{\prime }.
\end{equation*}%
In the language of the generalized method of moments (GMM) literature, the
summary statistics of which $\mathbf{\eta }(\mathbf{y})\ $is comprised
\textquotedblleft over-identify\textquotedblright\ $\mathbf{\theta }^{0}.$
In this case,\textbf{\ }[S2(2)] will be satisfied if the following equation
has a unique solution for all $\mathbf{\theta }^{i},\mathbf{\theta }^{0}$ $%
\in \Theta $: 
\begin{equation*}
0=\mathbf{b(\theta }^{0}\mathbf{)-b(\theta }^{i}\mathbf{)}=\left( 
\begin{array}{c}
1+(\theta _{1}^{0})^{2}+(\theta _{2}^{0})^{2} \\ 
\theta _{1}^{0}(1+\theta _{2}^{0}) \\ 
\theta _{2}^{0}%
\end{array}%
\right) -\left( 
\begin{array}{c}
1+(\theta _{1}^{i})^{2}+(\theta _{2}^{i})^{2} \\ 
\theta _{1}^{i}(1+\theta _{2}^{i}) \\ 
\theta _{2}^{i}%
\end{array}%
\right) .
\end{equation*}%
The additional (linear) restriction, $0=\theta _{2}^{0}-\theta _{2}^{i},$
ensures that the only value that satisfies $0=\mathbf{b(\theta }^{0}\mathbf{%
)-b(\theta }^{i}\mathbf{)}$ is now $\mathbf{\theta }^{0}=(\theta
_{1}^{0},\theta _{2}^{0})^{\prime }$, and consistency will be achieved as a
consequence.

Simply adding summary statistics to the ABC procedure is,\textit{\ however,}
not guaranteed to yield consistent inference: the chosen summary statistics
must be informative about the underlying parameters $\mathbf{\theta }$
governing the statistical properties of the structural model. To illustrate
this point, consider again the above example, but with the three-dimensional
vector summary statistic:%
\begin{equation*}
\mathbf{\eta }\left( \mathbf{y}\right) =(\eta _{0}\mathbf{(y),}\eta _{1}%
\mathbf{(y),}\eta _{3}\mathbf{(y)})^{\prime },
\end{equation*}%
where $\eta _{3}(\mathbf{y})=\frac{1}{T}\sum_{t=4}^{T}y_{t}y_{t-3}.$ \textit{%
Given the nature of the s}tructural model, $\eta _{3}(\mathbf{y})\overset{P}{%
\rightarrow }E[y_{t}y_{t-3}]=0$ and\textbf{\ }by construction $\eta _{3}(%
\mathbf{z}^{i})\overset{P}{\rightarrow }E[z_{t}^{i}z_{t-3}^{i}]=0$ for all $%
\mathbf{\theta }^{i}$. \textit{Hence, t}he summary statistic $\eta _{3}(%
\mathbf{y})$ yields no new information about $\mathbf{\theta }^{0}$ and does
not therefore produce a mapping $\mathbf{\theta }^{i}\longmapsto \mathbf{%
b(\theta }^{i}\mathbf{)}$ that is one-to-one.
\end{example}

\begin{figure}[h!]
	\centering
\includegraphics[width=7.0171in,height=3.4255in]{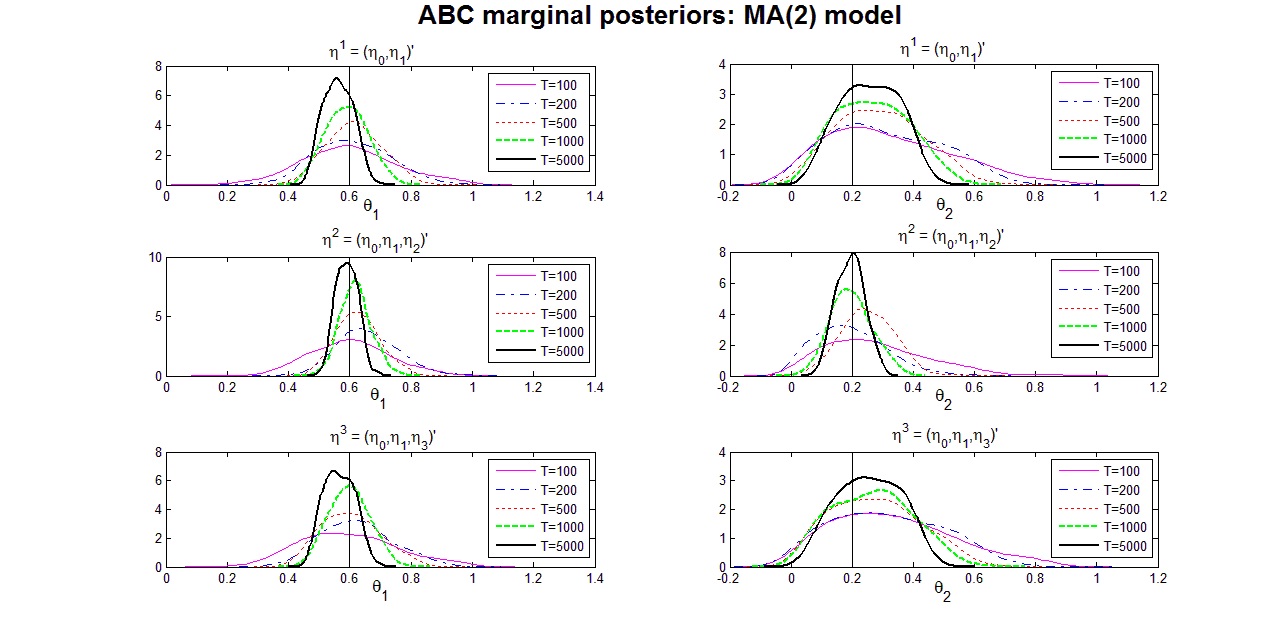}
\caption{ABC-based estimates of the
marginal posterior densities for the parameters of the MA(2) model, $\protect%
\theta _{1}$ and $\protect\theta _{2}$, with varying sample sizes. Top two
panels: summary statistic vector of $\mathbf{\protect\eta }^{1}\left( 
\mathbf{y}\right) =(\protect\eta _{0}\mathbf{(y),}\protect\eta _{1}\mathbf{%
(y)})^{\prime }\mathbf{;}$ Middle two panels: summary statistic vector of $%
\mathbf{\protect\eta }^{2}\left( \mathbf{y}\right) =(\protect\eta _{0}%
\mathbf{(y),}\protect\eta _{1}\mathbf{(y),}\protect\eta _{2}\mathbf{(y)})^{\prime }\mathbf{;}$ Bottom two panels: summary statistic vector of $%
\mathbf{\protect\eta }^{3}\left( \mathbf{y}\right) =(\protect\eta _{0}%
\mathbf{(y),}\protect\eta _{1}\mathbf{(y),}\protect\eta _{3}\mathbf{(y)}%
)^{\prime }\mathbf{.}$ The true parameter values are $\protect\theta %
_{1}^{0}=0.6$ and $\protect\theta _{2}^{0}=0.2$.}
\label{Fig1}
\end{figure}

We illustrate the theoretical results in Examples 2 and 3 graphically in
Figure \ref{Fig1}, denoting the three relevant vectors of summary statistics
as:%
\begin{eqnarray*}
\mathbf{\eta }^{1}\left( \mathbf{y}\right) &=&(\eta _{0}\mathbf{(y),}\eta
_{1}\mathbf{(y)})^{\prime }, \\
\mathbf{\eta }^{2}\left( \mathbf{y}\right) &=&(\eta _{0}\mathbf{(y),}\eta
_{1}\mathbf{(y),}\eta _{2}\mathbf{(y)})^{\prime }, \\
\mathbf{\eta }^{3}\left( \mathbf{y}\right) &=&(\eta _{0}\mathbf{(y),}\eta
_{1}\mathbf{(y),}\eta _{3}\mathbf{(y)})^{\prime }.
\end{eqnarray*}%
Using the true parameter vector $\mathbf{\theta }^{0}=(\theta
_{1}^{0},\theta _{2}^{0})^{\prime }=(0.6,0.2)^{\prime },$ a vector of
`observed' data, $\mathbf{y}=(y_{1},y_{2},...,y_{T})^{\prime }$ is
generated, for\textbf{\ } $T=100,$ $200,$ $500,$ $1,000$ and\textbf{\ }$%
5,000 $. For each given sample of size\textbf{\ }$T$, $p(\mathbf{\theta |y})$
is then estimated via the ABC method, using\ $N=50,000$ simulated draws from
uniform priors satisfying \eqref{const1}, and with the tolerance $%
\varepsilon _{j}$, $j=1,2,3,$ chosen so that only one-percent of the
simulated draws are accepted. {The top two panels} of Figure \ref{Fig1} {%
plot\ }$p_{\varepsilon _{1}}(\theta _{1}\mathbf{|\eta }^{1}\mathbf{(y)})$ and%
\textbf{\ }$p_{\varepsilon _{1}}(\theta _{2}\mathbf{|\eta }^{1}\mathbf{(y)})$
{respectively}, where the notation here indicates the kernel density
estimate of the relevant marginal density, conditional on\textbf{\ }$\mathbf{%
\eta }^{1}\mathbf{(y)}$,\textbf{\ }and as defined for the given $\varepsilon
_{1}.$ As the sample size increases both estimated marginals become more
concentrated, but not around the true values of $0.6$ and $0.2.$ In
contrast, the plots in {the two middle panels} demonstrate the consistency
that obtains when conditioning on\textbf{\ }$\mathbf{\eta }^{2}\mathbf{(y)}$,%
\textbf{\ }a result that is not replicated in the {two bottom panels}, in
which the three-dimensional conditioning vector is\textbf{\ }$\mathbf{\eta }%
^{3}\left( \mathbf{y}\right) .$\footnote{%
Whilst we have not pursued this in any formal way, the indications are that
in the two cases in which identification (of the true parameters) does not
obtain, the marginal posteriors are some form of mixture distribution, each
with a mode (or modes) that reflects (reflect) the location of the two pairs
of parameter values that satisfy (\ref{Truth}).}\textbf{\ }
\begin{description}
\item[Remark 5] The above example illustrates that adding additional summary
statistics to an ABC procedure may or may not aid researchers in obtaining
consistent inference. In particular, adding summary statistics will only be
helpful if the additional statistics contain information about the
parameters that is not accounted for by the summary statistics already used
in the analysis. Therefore, \textit{arbitrarily} adding summary statistics
will not necessarily yield valid inference. Moreover, and as was noted in
Section \ref{abc}, given that adding summary statistics hampers our ability
to accurately estimate the associated conditional density, adding summary
statistics to any initial ABC procedure should be embarked upon with care.

\item[Remark 6] It is also important to note that no link is to be expected
between the particular model at hand and the likelihood of Assumption
[S2(2)] being satisfied. As the above examples illustrate, it is the\textbf{%
\ }\textit{combination }of the model structure and the choice of summary
statistics that determines Bayesian consistency via ABC.
\end{description}

\section{Detecting Consistency\label{diag}}

\subsection{Preliminaries\label{prelim}}

Beyond understanding the theoretical conditions that must hold in order for
a particular set of summary statistics\ to yield valid inference, and noting
that in complex settings verifying the conditions of Theorem 1 will
typically not be possible via analytical means, it is useful to have some
way of ascertaining numerically whether those conditions actually hold in
any given case. To this end, we present a diagnostic tool that can be used
to determine if the estimated posterior obtained using a specific set of
summary statistics, say $\mathbf{\eta }(\mathbf{y})$, is Bayesian consistent.

The key insight to understanding the diagnostic procedure is that if the
true value $\mathbf{\theta }^{0}$ were known, we would only require a local
version of the identification condition (Assumption [S2(2)]); i.e., we would
only need to check that there existed no\textbf{\ }$\mathbf{\theta }^{\ast }$%
, with\textbf{\ }$\mathbf{\theta }^{\ast }\neq \mathbf{\theta }^{0}$,\textbf{%
\ }for which\textbf{\ }$\mathbf{b}(\mathbf{\theta }^{\ast })=\mathbf{b}(%
\mathbf{\theta }^{0}).$\textbf{\ }However, because\textbf{\ }$\mathbf{\theta 
}^{0}$\textbf{\ }is unknown, a sufficient condition to ensure that the above
holds is that the map\textbf{\ }$\mathbf{\theta \mapsto b(\theta )}$ is
one-to-one; i.e., that\textbf{\ }$\mathbf{b}(\mathbf{\theta }^{0})-$ $%
\mathbf{b}(\mathbf{\theta }^{\ast })=\mathbf{0}$ yields the unique solution%
\textbf{\ }$\mathbf{\theta }^{\ast }=\mathbf{\theta }^{0}$ for each and
every possible value of $\mathbf{\theta }^{0}.$ In this way, detecting
Bayesian consistency in ABC\ reduces to detecting satisfaction of the
one-to-one mapping assumption. The diagnostic procedure we propose seeks to
verify this condition, and hence the consistency of ABC posterior estimates,
in two stages: firstly, as it is applied to the observed data \textbf{$%
\mathbf{y}$ }(Section \ref{method 1}), and secondly, in terms of its
repeated application to data sets artificially generated from the assumed
true data generating process and across the feasible parameter space
(Section \ref{idenF2}).

The verification procedure exploits the following two facts: 1) under the
conditions of Theorem \ref{thm1}, the possible set of solutions for which $%
d\{\mathbf{\eta }(\mathbf{y}),\mathbf{\eta (z(\theta ))}\}=o_{P}(1)$ always
includes\textbf{\ }$\mathbf{\theta }=\mathbf{\theta }^{0}$; 2) if $d\{%
\mathbf{\eta (y),\eta (z(\theta ))}\}=o_{P}(1)$ uniquely at $\mathbf{\theta }%
=\mathbf{\theta }^{0}$, then an ABC procedure based on an augmented vector
of summary statistics, $\mathbf{\gamma (y)}=(\mathbf{\eta }(\mathbf{y}%
)^{\prime },\mathbf{g}(\mathbf{y})^{\prime })^{\prime }$, will yield a
posterior that is Bayesian consistent, so long as $\mathbf{g}(\cdot )$
satisfies conditions [S1] and [S2(1)] of Theorem \ref{thm1}. To state these
results more formally, assume $\mathbf{\eta }(\mathbf{y})$ (respectively, $%
\mathbf{g}(\mathbf{y})$) has a well-defined limit $\mathbf{b}_{1}(\mathbf{%
\theta }^{0})$ (respectively,\textbf{\ }$\mathbf{b}_{2}(\mathbf{\theta }%
^{0}) $) and denote the limit quantity of $\mathbf{\eta }(\mathbf{z}^{i})$
(respectively, $\mathbf{g}(\mathbf{z}^{i})$) as $\mathbf{b}_{1}(\mathbf{%
\theta }^{i})$ (respectively, $\mathbf{b}_{2}(\mathbf{\theta }^{i})$).

\begin{corollary}
\label{corr1} Given summary statistics $\mathbf{\gamma (y)}=(\mathbf{\eta }(%
\mathbf{y})^{\prime },\mathbf{g}(\mathbf{y})^{\prime })^{\prime }$, assume
that\textbf{\ }the following conditions\textbf{\ }are satisfied:

\begin{description}
\item[{[C0]}] The DGP for $\mathbf{y}$ is uniquely defined at $\mathbf{\theta 
}^{0}.$

\item[{[C1]}] For $\mathbf{b}(\mathbf{\theta }^{0})=(\mathbf{b}_{1}(\mathbf{%
\theta }^{0})^{\prime },\mathbf{b}_{2}(\mathbf{\theta }^{0})^{\prime
})^{\prime }$, we have $\Vert \mathbf{\gamma }(\mathbf{y})-\mathbf{b}(%
\mathbf{\theta }^{0})\Vert =o_{P}(1)$.

\item[{[C2]}] The map $\mathbf{\theta }^{i}\mapsto \mathbf{b}(\mathbf{\theta }%
^{i})$ is deterministic, continuous, exists for all\textbf{\ }$\mathbf{%
\theta }^{i}\in \mathbf{\Theta },$ and satisfies

\begin{description}
\item[{[C2(1)]}] $\sup_{\mathbf{\theta \in \Theta }}\Vert \mathbf{\gamma }(%
\mathbf{z}(\mathbf{\theta }))-\mathbf{b}(\mathbf{\theta })\Vert =o_{P}(1)$,

\item[{[C2(2)]}] $\mathbf{b}_{1}(\mathbf{\cdot })$ is one-to-one in $\mathbf{%
\theta }^{i}.$
\end{description}
\end{description}

\noindent%
If [C0]-[C2] and [P] are satisfied, then, for all $\delta >0,$ Pr$%
_{\varepsilon }(\mathbf{\theta }\notin \aleph _{\delta }(\mathbf{\theta }%
^{0})|\mathbf{\gamma (y))}\overset{P}{\rightarrow }0$ as $T\rightarrow
\infty $ and $\varepsilon \rightarrow 0$.
\end{corollary}

\subsection{Use of the Observed Data\label{method 1}}

To understand the implications of Corollary \ref{corr1}, consider the case
where we have already obtained $p_{\varepsilon _{1}}(\mathbf{\theta |\eta
(y))}$, for some tolerance $\varepsilon _{1},$ with $\mathbf{\eta (y)}$
based on a value of $T$ that is assumed to be large enough for large sample
behavior to be in evidence. Now, if we were to run ABC again using the joint
summary statistic $\mathbf{\gamma (y)}$, Corollary \ref{corr1} implies that
one of two things will happen: either the posterior $p_{\varepsilon _{2}}(%
\mathbf{\theta |\gamma (y)),}$ computed for some tolerance $\varepsilon
_{2}, $ will be located in a very similar position to $p_{\varepsilon _{1}}(%
\mathbf{\theta |\eta (y))}$, only potentially flatter or with a slightly
different shape, a consequence of the increased\textbf{\ }dimensionality,%
\footnote{%
Simulation evidence suggests that the increased flatness (or otherwise) of
the subsequent posterior estimates depends on the nature of the\textbf{\ }%
information about $\mathbf{\theta }^{0}$ contained in the additional summary
statistics.} or the high mass region of $p_{\varepsilon _{2}}(\mathbf{\theta
|\gamma (y))}$ will be located in a distinctly different part of the support
from that of $p_{\varepsilon _{1}}(\mathbf{\theta |\eta (y))}.$ We refer to
this latter event as one of $p_{\varepsilon _{2}}(\mathbf{\theta |\gamma (y))%
}$ \textquotedblleft jumping away\textquotedblright\ from\textbf{\ }$%
p_{\varepsilon _{1}}(\mathbf{\theta |\eta (y))}$ and, according to Corollary %
\ref{corr1}, see the occurrence of this event as evidence that the initial
summary statistics did \emph{not} yield a posterior that is Bayesian
consistent. If, on the other hand, the addition of $\mathbf{g}(\mathbf{y})$
does not cause the mass of $p_{\varepsilon _{2}}(\mathbf{\theta |\gamma (y))}
$ to jump in relation to $p_{\varepsilon _{1}}(\mathbf{\theta |\eta (y))}$,
then this suggests that the initial choice of summary statistics, $\mathbf{%
\eta (y),}$\textbf{\ }may have yielded valid inference. The use of the word
`may' reflects the fact that there is no guarantee possible, via use of the
observed data alone, that consistency has been achieved, since there is no
guarantee \textit{a priori} that $d\{\mathbf{\eta (y),\eta (z(\theta ))}%
\}=o_{P}(1)$ has a\textbf{\ }\textit{unique} solution $\mathbf{\theta }=%
\mathbf{\theta }^{0}$. It is this point that is addressed in next subsection.
\begin{figure}[h!]
	\centering
\includegraphics[width=7.3068in, height=3.2128in]{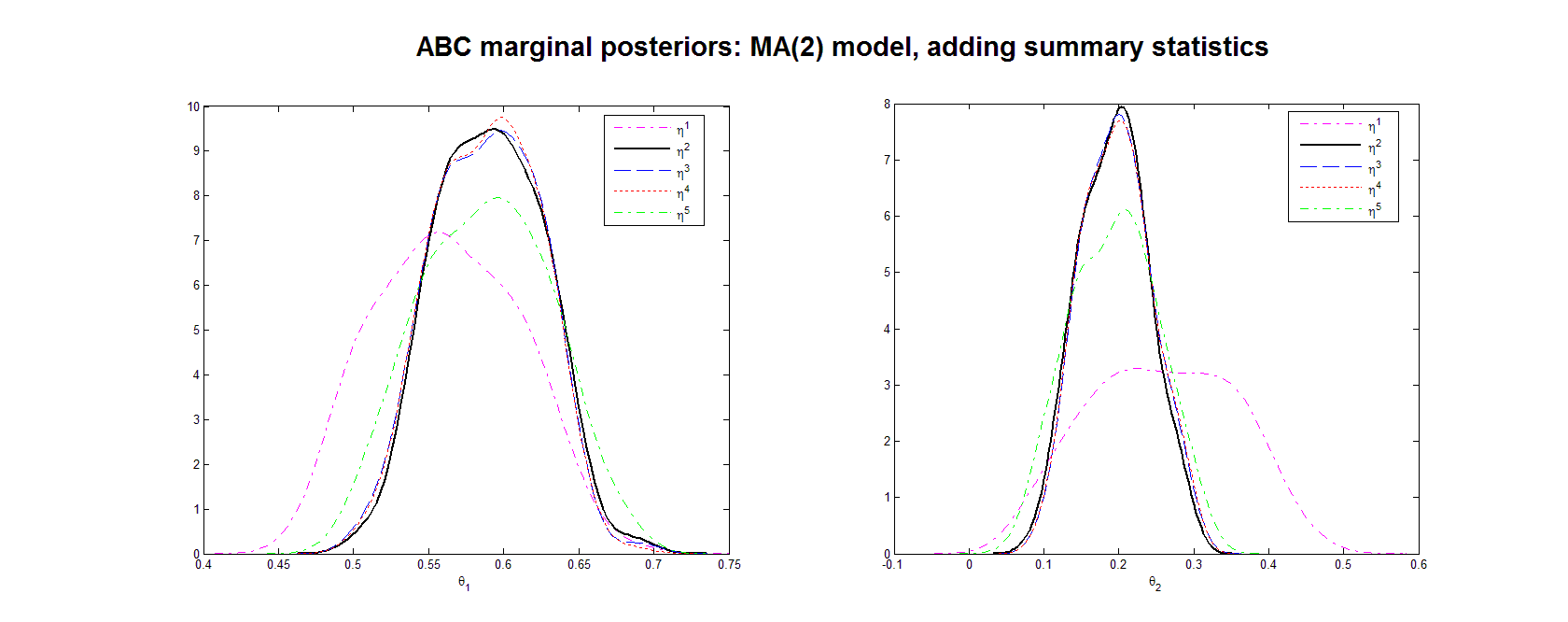}
\caption{ABC-based estimates of the
marginal posterior densities for the parameters of the MA(2) model, $\protect%
\theta _{1}$ and $\protect\theta _{2}$, with $T=5000$. The key for both
panels indicates the\textbf{\ }summary statistic vector used: $\mathbf{%
\protect\eta }^{1}\left( \mathbf{y}\right) =(\protect\eta _{0}\mathbf{(y),}%
\protect\eta _{1}\mathbf{(y)})^{\prime }\mathbf{;}$ $\mathbf{\protect\eta }%
^{2}\left( \mathbf{y}\right) =(\mathbf{\protect\eta }^{1}(\mathbf{y}%
)^{\prime },\protect\eta _{2}(\mathbf{y}))^{\prime }\mathbf{;}$ $\mathbf{%
\protect\eta }^{3}(\mathbf{y})=(\mathbf{\protect\eta }^{2}(\mathbf{y}%
)^{\prime },\protect\eta _{3}(\mathbf{y}))^{\prime }\mathbf{;}$ $\mathbf{%
\protect\eta }^{4}(\mathbf{y})=(\mathbf{\protect\eta }^{3}(\mathbf{y}%
)^{\prime },\frac{1}{T}\sum_{t=1}^{T}y_{t})^{\prime };$ $\mathbf{\protect%
\eta }^{5}\mathbf{(y)}=(\mathbf{\protect\eta }^{4}(\mathbf{y})^{\prime },%
\frac{1}{T}\sum_{t=1}^{T}y_{t}^{3})^{\prime }.$ The statistics $\mathbf{%
\protect\eta }^{2}(\mathbf{y})$\ to $\mathbf{\protect\eta }^{5}(\mathbf{y})$%
\ yield Bayesian consistency. The true parameter values are $\protect\theta %
_{1}^{0}=0.6$ and $\protect\theta _{2}^{0}=0.2$.}
\label{Fig2}
\end{figure}

Meanwhile, we illustrate this preliminary diagnostic exercise via the MA(2)
model, in which case (from Example 3) we have an analytical result that
establishes that consistency for the true $\mathbf{\theta }^{0}=(\theta
_{1}^{0},\theta _{2}^{0})^{\prime }=(.6,.2)^{\prime }$ is\ achieved via a
particular choice of summary statistics. We adopt five different choices of
summary statistics for use in the illustration:%
\begin{eqnarray*}
\mathbf{\eta }^{1}(\mathbf{y}) &=&(\eta _{0}(\mathbf{y}),\eta _{1}(\mathbf{y}%
))^{\prime } \\
\mathbf{\eta }^{2}(\mathbf{y}) &=&(\mathbf{\eta }^{1}(\mathbf{y})^{\prime
},\eta _{2}(\mathbf{y}))^{\prime } \\
\mathbf{\eta }^{3}(\mathbf{y}) &=&(\mathbf{\eta }^{2}(\mathbf{y})^{\prime
},\eta _{3}(\mathbf{y}))^{\prime } \\
\mathbf{\eta }^{4}(\mathbf{y}) &=&(\mathbf{\eta }^{3}(\mathbf{y})^{\prime },%
\frac{1}{T}\sum_{t=1}^{T}y_{t})^{\prime } \\
\mathbf{\eta }^{5}\mathbf{(y)} &=&(\mathbf{\eta }^{4}(\mathbf{y})^{\prime },%
\frac{1}{T}\sum_{t=1}^{T}y_{t}^{3})^{\prime },
\end{eqnarray*}%
where $\eta _{j}(\mathbf{y})=\frac{1}{T}\sum_{t=1+j}^{T}y_{t}y_{t-j}\,,$ for 
$j=0,1,2,3.$ We set the sample size to\textbf{\ }$T=5,000$,\textbf{\ }%
consider\textbf{\ }$N=50,000$\textbf{\ }simulations and set the tolerance%
\textbf{\ }$\varepsilon _{j},$\textbf{\ }$j=1,...,5$, so that we retain
one-percent of the simulated draws for each choice of summary statistics.
From our previous theoretical analysis we know that $\mathbf{\eta }^{1}(%
\mathbf{y})$ will not yield an estimated posterior, $p_{\varepsilon _{1}}(%
\mathbf{\theta }|\mathbf{\eta }^{1}(\mathbf{y}))$, that is Bayesian
consistent, while the remaining sets will yield posteriors that \textit{are}
Bayesian consistent, due to the inclusion of\textbf{\ }$\eta _{2}\mathbf{(y).%
}$ Therefore, after adding $\eta _{2}\mathbf{(y)}$ to our initial choice of
summary statistics, $\mathbf{\eta }^{1}(\mathbf{y}),$ the estimated
posterior $p_{\varepsilon _{2}}(\mathbf{\theta }|\mathbf{\eta }^{2}(\mathbf{y%
}))$ should be centered around the true values, or thereabouts (given the
still finite value of\textbf{\ }$T$); that is, the main mass of the posterior%
\textbf{\ }computed using\textbf{\ }$\mathbf{\eta }^{2}(\mathbf{y})$ should
\textquotedblleft jump\textquotedblright\ away from the main mass of the
posterior computed using\textbf{\ }$\mathbf{\eta }^{1}(\mathbf{y})$.
Subsequently, the posteriors based on summary statistics $\mathbf{\eta }^{3}%
\mathbf{(y)},\mathbf{\eta }^{4}\mathbf{(y)}$ and\textbf{\ }$\mathbf{\eta }%
^{5}\mathbf{(y)}$ should not move much, if at all, in relation to $%
p_{\varepsilon _{2}}(\mathbf{\theta }|\mathbf{\eta }^{2}(\mathbf{y}))$ but
may possibly become flatter,\textbf{\ }and possibly change shape, with each
additional summary statistic. Figure \ref{Fig2} illustrates these points
exactly. The estimated posterior $p_{\varepsilon _{2}}(\mathbf{\theta }|%
\mathbf{\eta }^{2}\mathbf{(y)})$ is seen to shift substantially in relation
to the estimated posterior $p_{\varepsilon _{1}}(\mathbf{\theta }|\mathbf{%
\eta }^{1}\mathbf{(y)})$. In turn, adding $\eta _{3}(\mathbf{y})$ and $\frac{%
1}{T}\sum_{t=1}^{T}y_{t}$ to $\mathbf{\eta }^{2}\mathbf{(y)}$ causes minimal
change, and certainly no discernible change in location. The location of the
high mass point is preserved by the subsequent addition of\textbf{\ }$\frac{1%
}{T}\sum_{t=1}^{T}y_{t}^{3}$; however at this point the dimension of the
full statistic\textbf{\ }$\mathbf{\eta }^{5}\mathbf{(y)}$\textbf{\ }appears
to cut in, with the accuracy of the kernel density estimation adversely
affected.\footnote{%
Results for $T=1,000$ and $T=10,000$ were also considered. The resulting
plots paint a similar picture and hence have not been included for brevity.}%
\begin{figure}[h!]
	\centering
\includegraphics[width=7.3068in, height=3.2128in]{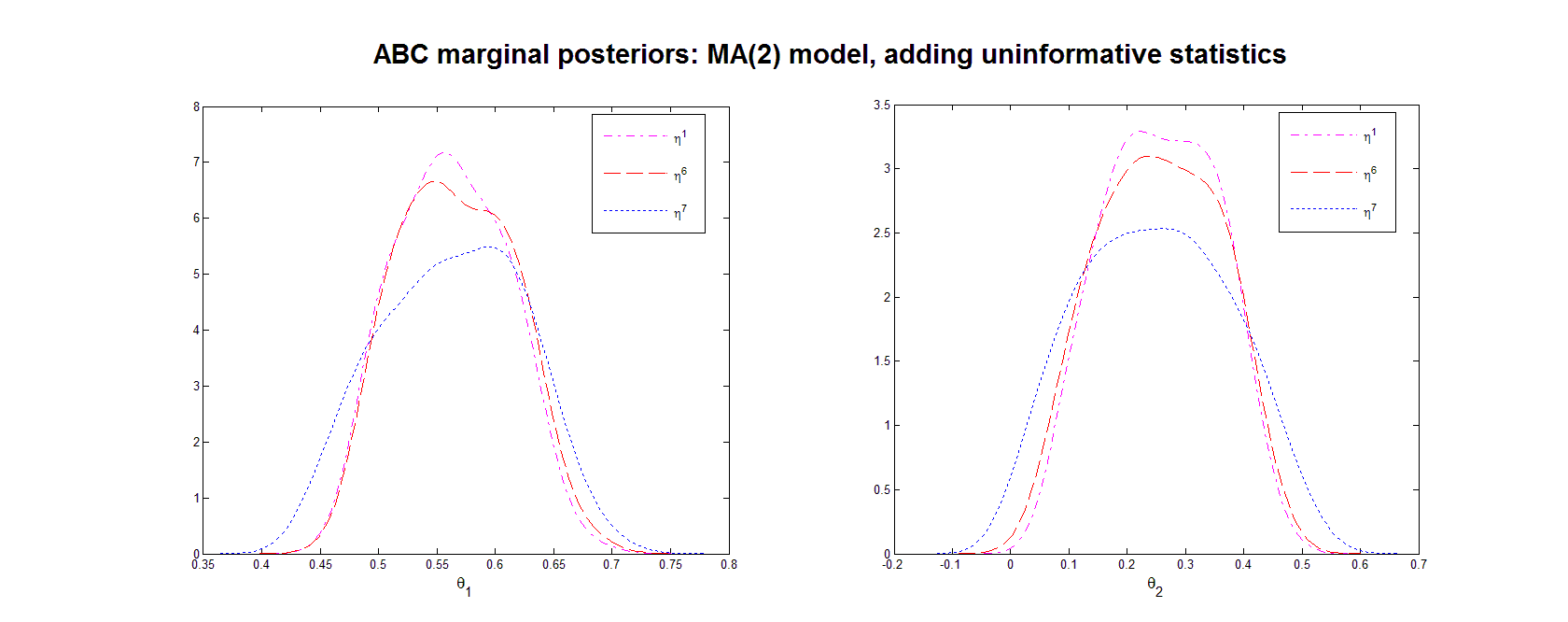}
\caption{ABC-based estimates of the
marginal posterior densities for the parameters of the MA(2) model, $\protect%
\theta _{1}$ and $\protect\theta _{2}$, with $T=5000$. The key for both
panels indicates the\textbf{\ }summary statistic vector used: $\mathbf{%
\protect\eta }^{1}\left( \mathbf{y}\right) =(\protect\eta _{0}\mathbf{(y),}%
\protect\eta _{1}\mathbf{(y)})^{\prime }\mathbf{;}$ $\mathbf{\protect\eta }%
^{6}\mathbf{(y)}=(\mathbf{\protect\eta }^{1}\mathbf{(y)},\protect\eta _{3}(%
\mathbf{y}))^{\prime }\mathbf{;}$ $\mathbf{\protect\eta }^{7}\mathbf{(y)}=(%
\mathbf{\protect\eta }^{6}\mathbf{(y)},\frac{1}{T}\sum_{t=1}^{T}y_{t}^{3})^{%
\prime }.$ All three summary statistics do not yield Bayesian consistency.%
\textbf{\ }The true parameter values are $\protect\theta _{1}^{0}=0.6$ and $%
\protect\theta _{2}^{0}=0.2$. }
\label{Fig4}
\end{figure}
\begin{figure}[h!]
	\centering
\includegraphics[width=7.3068in, height=3.2128in]{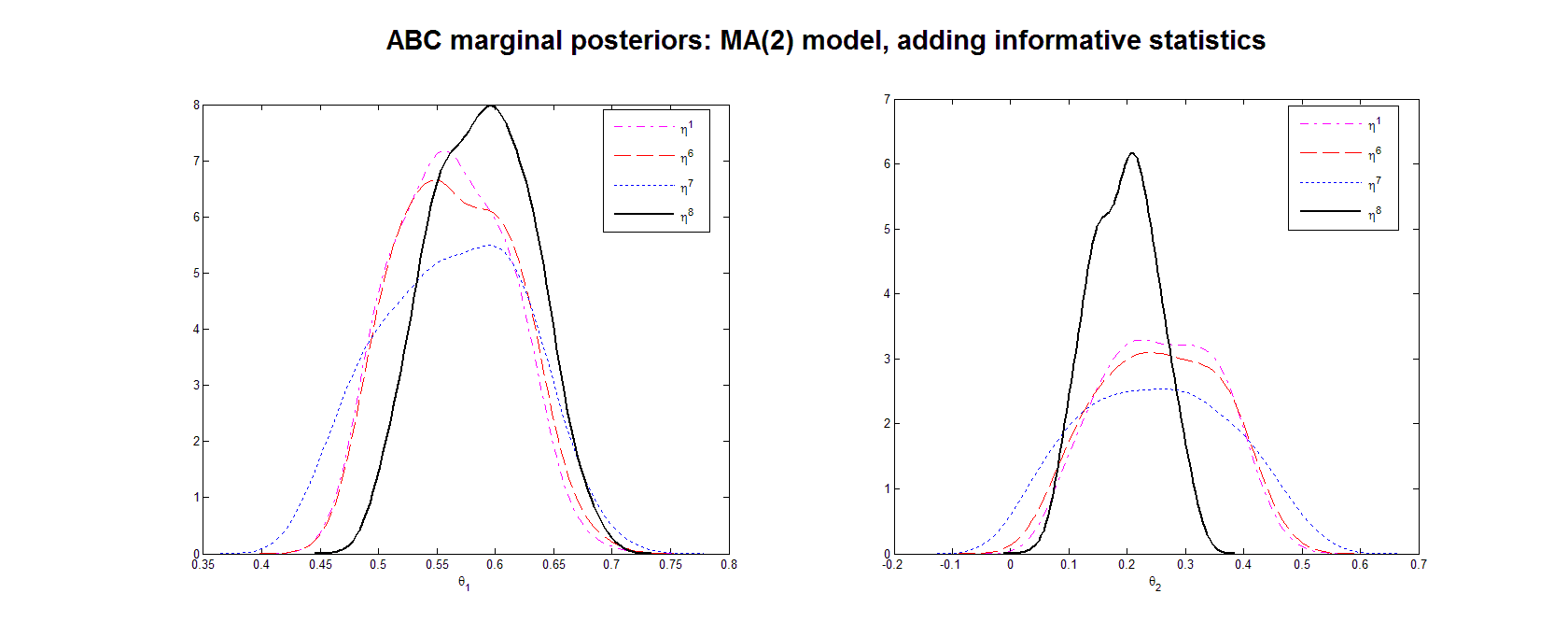}
\caption{ABC-based estimates of the
marginal posterior densities for the parameters of the MA(2) model, $\protect%
\theta _{1}$ and $\protect\theta _{2}$, with $T=5000$. The key for both
panels indicates the\textbf{\ }summary statistic vector used: $\mathbf{%
\protect\eta }^{1}\left( \mathbf{y}\right) =(%
\protect\begin{array}{cc}
\protect\eta _{0}\mathbf{(y),} & \protect\eta _{1}\mathbf{(y)}%
\protect\end{array}%
)^{\prime }\mathbf{;}$ $\mathbf{\protect\eta }^{6}\mathbf{(y)}=(\mathbf{%
\protect\eta }^{1}\mathbf{(y)},\protect\eta _{3}(\mathbf{y}))^{\prime }%
\mathbf{;}$ $\mathbf{\protect\eta }^{7}\mathbf{(y)}=(\mathbf{\protect\eta }%
^{6}\mathbf{(y)},\frac{1}{T}\sum_{t=1}^{T}y_{t}^{3})^{\prime };$ $\mathbf{%
\protect\eta }^{8}\mathbf{(y)}=(\mathbf{\protect\eta }^{7}\mathbf{(y)},%
\protect\eta _{2}(\mathbf{y}))^{\prime }.$ The first three summary
statistics do not yield\textbf{\ }Bayesian\textbf{\ }consistency.\textbf{\ }%
The fourth is associated with consistency and the marginal posterior
estimates are shown to differ markedly from the first three as a
consequence. The true parameter values are $\protect\theta _{1}^{0}=0.6$ and 
$\protect\theta _{2}^{0}=0.2$.}
\label{Fig5}
\end{figure}

Let us now consider a similar exercise with summary statistics $\mathbf{\eta 
}^{1}\mathbf{(y)}=(\eta _{0}(\mathbf{y}),\eta _{1}(\mathbf{y}))^{\prime },%
\mathbf{\eta }^{6}\mathbf{(y)}=(\mathbf{\eta }^{1}\mathbf{(y)},\eta _{3}(%
\mathbf{y}))^{\prime },\mathbf{\eta }^{7}\mathbf{(y)}=(\mathbf{\eta }^{6}%
\mathbf{(y)},\frac{1}{T}\sum_{t=1}^{T}y_{t}^{3})^{\prime }$ {and }$\mathbf{%
\eta }^{8}\mathbf{(y)}=(\mathbf{\eta }^{7}\mathbf{(y)},\eta _{2}(\mathbf{y}%
))^{\prime }$, where we deliberately use different notation to distinguish
these statistics from those used in the illustration above. In this
particular setup the only set of summary statistics that will yield
consistent inference is $\mathbf{\eta }^{8}\mathbf{(y)}$, and the aim of the
exercise is to illustrate the\textit{\ differential}\textbf{\ }impact of
adding non-informative and informative summary statistics to an initial set
that does not yield identification. First, consider Figure \ref{Fig4}, which
plots posteriors based only on\textbf{\ }$\mathbf{\eta }^{1}\mathbf{(y),\eta 
}^{6}\mathbf{(y)}$ and\textbf{\ }$\mathbf{\eta }^{7}\mathbf{(y).}$\textbf{\ }%
Adding $\eta _{3}(\mathbf{y})$ to $\mathbf{\eta }^{1}\mathbf{(y)}$ to produce%
\textbf{\ }$\mathbf{\eta }^{6}\mathbf{(y)}$ (which we know does not yield
identification) causes the estimated posterior $p_{\varepsilon _{6}}(\mathbf{%
\theta }|\mathbf{\eta }^{6}(\mathbf{y}))$ to flatten out compared to $%
p_{\varepsilon _{1}}(\mathbf{\theta }|\mathbf{\eta }^{1}(\mathbf{y}))$, and
to shift slightly. Now, adding the statistic $\frac{1}{T}%
\sum_{t=1}^{T}y_{t}^{3}$ to $\mathbf{\eta }^{6}\mathbf{(y)}$, knowing as we
do that this statistic will also not aid in identification\footnote{%
For the MA(2) model in (\ref{MA2}), with $e_{t}\sim i.i.d.N(0,1),$ $%
E(y_{t}^{3})=E(e_{t}+\theta _{1}e_{t-1}+\theta _{2}e_{t-1})^{3}$ is composed
of four different types of moments: $%
E(e_{t-k}^{3}),E(e_{t-k}^{2}e_{t-j}),E(e_{t-k}e_{t-j})$ and $%
E(e_{t-k}e_{t-j}e_{t-l})$, for $l\neq k\neq j$, which are all zero for any
value of $\mathbf{\theta }=(\theta _{1},\theta _{2})^{\prime }$.}, the
posterior $p_{\varepsilon _{7}}(\mathbf{\theta }|\mathbf{\eta }^{7}(\mathbf{y%
}))$ becomes even flatter (reflecting the increased dimension) and continues
to shift away from the posterior mode of both previously estimated
posteriors, a clear indication that we did not yet have a valid set of
summary statistics on the previous rounds.

In Figure \ref{Fig5} we then superimpose on these three plots the estimated
marginal posterior based on\textbf{\ }$\mathbf{\eta }^{8}\mathbf{(y)}$,
where we know that the combination of $\mathbf{\eta }^{7}\mathbf{(y)}$ and $%
\eta _{2}(\mathbf{y})$ (which defines\textbf{\ }$\mathbf{\eta }^{8}\mathbf{%
(y)}$)\textbf{\ }contains sufficient information for the parameters to now
be identified. The change in the estimated posterior $p_{\varepsilon _{8}}(%
\mathbf{\theta }|\mathbf{\eta }^{8}(\mathbf{y}))$, relative to the existing
three, is marked, with a clear peak observed around the true values, $\theta
_{1}^{0}=0.6$ and $\theta _{2}^{0}=0.2$.\textbf{\ }Subsequent\textit{\ }%
additions of statistics to this set will, along the lines illustrated in
Figure \ref{Fig2}, produce posteriors that now remain reasonably fixed at
the same modal value and that vary only in terms of dispersion, if at all.%
\footnote{%
The experiments in Section \ref{method 1} are conducted using raw distances
(no component scaling). However, the results were also conducted using
distances scaled by the sample covariance matrix of the summary statistics,
and with individual elements scaled by their simulated variance. Results
based on these alternative scaling measures are not qualitatively different
from those presented herein. The results are available from the authors upon
request.}

The results of this section are summed up in the following remarks:

\begin{description}
\item[Remark 7] Corollary \ref{corr1} says that if ABC based on summary
statistics $\mathbf{\eta }(\mathbf{y})$ yields consistent inference, adding
more information, in the form of additional summary statistics, will never
invalidate the inference. Typically, adding further statistics can `dull'
the inference, in terms of producing a more dispersed posterior, {or a
posterior} with slightly different shape; but it will not shift the mode.
Hence, repeated augmentation of an initial choice of statistics, whereby the
mode of the estimated posterior eventually `settles' at a particular
location, should instill some confidence in the mind of the investigator
that consistency may have been achieved.

\item[Remark 8] Whilst the impact of adding non-informative statistics to an
initially non-informative set is likely to be problem-specific, we speculate
that small and continual changes in both location and dispersion are
indicative that a sufficiently informative set of summary statistics has not
yet been located. A more substantial shift at some point, followed by a lack
of change in the location at least, with the subsequent additions of
statistics, is indicative that identification and, hence, consistency, may
have been achieved. As flagged above, however, important caveats pertaining
to this statement are pursued in the following section.

\item[Remark 9] The above procedure has a similar flavor to the stepwise
search algorithm proposed in Joyce and Marjoram (2008). Despite this
apparent similarity however, the two procedures differ in terms of their
details, as well as having very different objectives. To wit, whilst the
approach outlined above is concerned with obtaining\ a vector of summary
statistics that yield consistent inference, that of Joyce and Marjoram is
concerned with obtaining a vector of summary statistics that is as
informative as possible (or `approximately sufficient' to use their
terminology) for any given $T.$
\end{description}

\subsection{Use of Repeated Simulation\label{idenF2}}

We have demonstrated how the numerical procedure proposed above can
determine with some certainty whether or not\textbf{\ }$p_{\varepsilon }%
\mathbf{(\theta |\eta (y))}$\textbf{\ }- for some choice of\textbf{\ }$%
\mathbf{\eta (y)}$ - is concentrating at\textbf{\ }$\mathbf{\theta }^{0}$,%
\textbf{\ }in the artifical scenario in which $\mathbf{\theta }^{0}$\ is
known. In practice of course, the true value $\mathbf{\theta }^{0}$\ is
unknown, and the proposed method is not capable of distinguishing between%
\textbf{\ }$p_{\varepsilon }(\mathbf{\theta |\eta (y)})$\textbf{\ }%
concentrating at\textbf{\ }$\mathbf{\theta }^{0}$\textbf{\ }and\textbf{\ }$%
p_{\varepsilon }(\mathbf{\theta |\eta (y)})$\textbf{\ }concentrating at some
other value\textbf{\ }$\mathbf{\theta }^{\ast }\neq \mathbf{\theta }^{0}$,
satisfying\textbf{\ }$\mathbf{b(\theta }^{\ast }\mathbf{)}=\mathbf{b(\theta }%
^{0}\mathbf{)}$. However, if the binding function is one-to-one, perverse
situations such as the above can be ruled out.

For a fixed (vector) summary statistic,\textbf{\ }$\mathbf{\eta (y)}$, it
turns out that verifying whether or not the binding function is one-to-one
is, in principle,\textbf{\ }possible. To understand how we can verify this
condition, first recall that $\mathbf{b(\theta )}$ is simply the limit, as $%
T\rightarrow \infty $, of the simulated summary statistics$\ \mathbf{\eta
(z(\theta ))}$, and note that because $\mathbf{z(\theta )}$ is simulated
from the structural model, $\mathbf{z(\theta )}$ is no longer restricted to
be of the same length as the observed sample $\mathbf{y}$. From these facts,
we see that our ability to obtain $\mathbf{b(\theta )}$ is limited only by
computational power and time; i.e., we are limited only by our ability to
simulate (very) long trajectories for $\mathbf{z(\theta )}$. In addition,
the entire map $\mathbf{\theta }\mapsto \mathbf{b(\theta )}$ can be obtained
simply by simulating long trajectories of $\mathbf{z(\theta )}$, forming $%
\mathbf{\eta (z(\theta ))}$, and repeating the exercise at every $\mathbf{%
\theta }\in \mathbf{\Theta }$.\textbf{\ }Therefore, with enough computing
power (and time), it is theoretically possible to verify whether or not $%
\mathbf{\theta }\mapsto \mathbf{b(\theta )}$ is one-to-one.

While the above logic demonstrates that it is theoretically possible to
verify the one-to-one condition, it is not practically possible as this
approach (technically) requires simulating an infinite number of infinite
series. However, when the data is stationary and the parameter space
relatively small, we can approximately check this condition through the
following steps: 
\begin{algorithm}
\caption{One-to-one verification}\label{one_ABC}
\begin{algorithmic}[1]
\State Use the empirical procedure in Section \ref{method 1} to identify
a (vector) summary statistic of interest, hereafter denoted $\mathbf{\eta
(\cdot ).}$
\State Choose $K^{\ast }$ distinct parameter values with which to
simulate data from the structural model, call them $\mathbf{\theta }%
^{0,1},...,\mathbf{\theta }^{0,K^{\ast }}$. Choose a large integer $T^{\ast
}>>T>>0$.
\State Simulate $\widetilde{{\mathbf{z}}}^{k}=(z_{1}(\mathbf{\theta }%
^{0,k}),...,z_{T^{\ast }}(\mathbf{\theta }^{0,k}))^{\prime }$ and form the
series $\left\{ \mathbf{\eta }(\widetilde{{\mathbf{z}}}^{k})\right\}
_{k=1}^{K^{\ast }}$; $\left\{ \mathbf{\eta }(\widetilde{{\mathbf{z}}}%
^{k})\right\} _{k=1}^{K^{\ast }}$ constitutes a discrete approximation to $%
\mathbf{\theta }\mapsto \mathbf{b(\theta )}$
\State Determine whether or not $\left\{ \mathbf{\eta }(\widetilde{{%
\mathbf{z}}}^{k})\right\} _{k=1}^{K^{\ast }}$ contains $K^{\ast }$ unique
elements.
\end{algorithmic}
\end{algorithm}\ \newline
For $K^{\ast }$ and $T^{\ast }$ large enough, if $\mathbf{\eta }(\cdot )$
satisfies Step (4), \textit{and} if the estimated $p_{\varepsilon }(\mathbf{%
\theta }|\mathbf{\eta }(\mathbf{y}))$, as based on the observed data,\textbf{%
\ }$\mathbf{y}$, is also\textbf{\ }collapsing toward some point, one should
conclude in favor of consistency. {Of course}, this process leaves much left
unspecified, with the most critical issues being how to span the parameter
space, how to selected the set of pre-specified statistics, and the order in
which they are to be explored, plus the manner in which degeneracy of the
estimated posterior is tested for as $T$ is allowed to increase. However,
providing guidelines for and proving the theoretical properties of any such
search procedure {would require} several layers of formalization and the
introduction of new terms and concepts that would detract from the current
message of the paper. Hence, at this stage we simply emphasize that a
completely satisfactory assessment of consistency would appear to require
both the use of the observed data and repeated application of data from the
assumed process; and suggest that the sort of exercise we are proposing
here, albeit informal, is a sensible one to pursue.

\section{Criterion Functions based on an Auxiliary Model\label{crit}}

\subsection{Consistency of Auxiliary Model-based ABC}

At a minimum, implementation of ABC requires some means of generating
summary statistics $\mathbf{\eta }(\mathbf{y})$ that are \textquotedblleft
informative\textquotedblright\ about the unknown parameters of the
underlying structural model; whereby \textquotedblleft
informative\textquotedblright\ it is generally meant that the summary
statistics are a useful way of characterizing the information contained in
the observed data. However, the previous sections demonstrate that care must
be taken to ensure that the chosen summary statistics yield consistent
inference.

An alternative way of obtaining informative summary statistics is through
the use of an auxiliary model that depends on parameters\textbf{\ }$\mathbf{%
\beta }\in \mathbf{B}\subset 
\mathbb{R}
^{d_{\beta }}$, where $d_{\beta }\geq p$=dim$(\mathbf{\Theta })$,\ and for
which the likelihood function of the auxiliary model, denoted by $L(\mathbf{y%
};\mathbf{\beta })$, is known in closed form.\textbf{\ }Given a simple
auxiliary likelihood $L(\mathbf{y};\mathbf{\beta })$, a growing literature
suggests using summary statistics generated from\ $L(\mathbf{y};\mathbf{%
\beta })$; for example, one can choose $\mathbf{\eta }(\mathbf{y})=\mathbf{%
\widehat{\beta }(y)},$ where $\mathbf{\widehat{\beta }(y)}=\arg \max_{\beta
\in \mathbf{B}}L(\mathbf{y};\mathbf{\beta })$,\textbf{\ }or $\mathbf{\eta }(%
\mathbf{y})$ equivalent to the vector score of $L(\mathbf{y};\mathbf{\beta }%
) $ evaluated at $\mathbf{\widehat{\beta }(y)}$. However, by its very nature
the auxiliary model, and by proxy the summary statistics derived from $L(%
\mathbf{y};\mathbf{\beta })$, is (are) likely to describe only \textit{%
certain} salient features of the underlying structural model. In particular,
there is generally no reason to believe that the auxiliary model should
\textquotedblleft nest\textquotedblright\ the true structural model in some
well-defined sense. Indeed, if it does so then this suggests either that the
structural model itself is tractable - hence excluding the need for ABC - or
that the nesting model is highly parameterized, thereby inducing a $\mathbf{%
\eta }(\mathbf{\cdot })$ of high dimension and the associated problems for
accuracy.

Given then that a typical auxiliary model is capable of representing only
certain salient features of the DGP, there is nothing particularly special
about choosing the auxiliary \textit{likelihood function }to generate
summary statistics for use within ABC. Moreover, in many cases a realistic
auxiliary model may yield a likelihood function that is itself too
complicated for ABC, from a purely computational standpoint, whilst an
alternative criterion function, based on the same auxiliary model, may yield
computationally simpler summary statistics. For example, alternative
criterion functions - other than an auxiliary likelihood - that could be
used inside an ABC algorithm include: sums of squared errors, least absolute
deviations, and even quadratic functions of sample moments (conditional and
unconditional) from an auxiliary model, with the latter used to define an
MSM-type of approach, but with moments of the auxiliary rather than the true
model defining the selection mechanism.

However, as in the previous section, conditions need to be placed on the
relevant criterion function to ensure the resultant ABC procedure yields
consistent inference. This is the content of Theorem\textbf{\ }\ref{thm2}.%
\textbf{\ }Begin by defining a sample criterion function based on observed
data $\mathbf{y}$\textbf{\ }(respectively, simulated data\textbf{\ }$\mathbf{%
z}^{i}=\mathbf{z(\theta }^{i}\mathbf{)})$\textbf{\ }$Q(\mathbf{y};\mathbf{%
\beta })$ (respectively, $Q(\mathbf{z}^{i};\mathbf{\beta })$) and define $%
\mathbf{\widehat{\beta }(y)}$ (respectively, $\mathbf{\widehat{\beta }}(%
\mathbf{z}^{i})$) as the minimizer of $Q(\mathbf{y};\mathbf{\beta })$
(respectively, $Q(\mathbf{z}^{i};\mathbf{\beta })$). For a particular choice
of $Q(\cdot ;\mathbf{\beta })$ an ABC algorithm could be based on the
summary statistics $\mathbf{\eta (y)=\widehat{\beta }(y)}$, $\mathbf{\eta }(%
\mathbf{z}^{i})=\mathbf{\widehat{\beta }}(\mathbf{z}^{i})$.

The above intuition yields ABC Algorithm \ref{ABC_aux} based on generic
criterion\textbf{\ $Q(.;\mathbf{\beta })$}:

\begin{algorithm}
\caption{ABC algorithm: auxiliary criterion function}\label{ABC_aux}
\begin{algorithmic}[1]
\State Obtain $\widehat{%
\mathbf{\beta }}\mathbf{(y)=\arg \min_{\beta \in \mathbf{B}}}Q(\mathbf{y};%
\mathbf{\beta })$,
\State Simulate $\mathbf{\theta }^{i}$, $i=1,2,...,N$, from $p(%
\mathbf{\theta })$
\State Simulate $\mathbf{z}%
^{i}=(z_{1}^{i},z_{2}^{i},...,z_{T}^{i})^{\prime }$, $i=1,2,...,N$, from the
likelihood, $p(\mathbf{.|\theta }^{i})$
\State Select $\mathbf{\theta }^{i}$ such that:%
\begin{equation}
d\{\mathbf{\eta }(\mathbf{y}),\mathbf{\eta }(\mathbf{z}^{i})\}=d\{\mathbf{%
\widehat{\beta }(y)},\mathbf{\widehat{\beta }}(\mathbf{z}^{i})\}\leq
\varepsilon ,  \label{general_distance}
\end{equation}%
where $d\{\cdot
,\cdot \}$ is a distance function (or metric), and the tolerance level $%
\varepsilon $ is chosen as small as the computing budget allows.
\end{algorithmic}
\end{algorithm}

Denote the posterior obtained from the above algorithm as $p_{\varepsilon
}^{Q}(\mathbf{\theta }|\mathbf{\eta }(\mathbf{y}))$. The following result
gives conditions under which $\text{Pr}_{\varepsilon }^{Q}(\mathbf{\theta
\notin \aleph }_{\delta }\mathbf{(\theta }^{0})|\mathbf{\eta }(\mathbf{y}))%
\overset{P}{\rightarrow }0$ as $T\rightarrow \infty $ and $\varepsilon
\rightarrow 0$.

\begin{theorem}
\label{thm2} For an auxiliary model with parameters $\mathbf{\beta }\in 
\mathbf{B}$, $\mathbf{B}\subset \mathbb{R} ^{d_{\beta }}$ compact, assume
that the following are satisfied:

\begin{description}
\item[{[G1]}] There exists a deterministic limit criterion function $%
Q_{\infty }(\mathbf{\theta }^{i};\mathbf{\beta })$ such that

\begin{description}
\item[{[G1(1)]}] $Q_{\infty }(\mathbf{\theta }^{i}\mathbf{;\beta })$ is
continuous as a function of $\mathbf{\beta }$, uniformly in $\mathbf{\theta }%
^{i}$.

\item[{[G1(2)]}] $\sup_{\beta \in \mathbf{B}}\left\vert Q(\mathbf{y};\mathbf{%
\beta })-Q_{\infty }(\mathbf{\theta }^{0}\mathbf{;\beta })\right\vert
=o_{P}(1)$ and $\ \sup_{\mathbf{\theta \in \Theta },\mathbf{\beta }\in 
\mathbf{B}}\left\vert Q(\mathbf{z(\theta )};\mathbf{\beta })-Q_{\infty }(%
\mathbf{\theta ;\beta })\right\vert =o_{P}(1)$
\end{description}

\item[{[G2]}] $Q_{\infty }(\mathbf{\theta }^{i}\mathbf{;\beta })$ has a
unique minimum $\mathbf{b}(\mathbf{\theta }^{i})$ for all $\mathbf{\theta }%
^{i}\in \mathbf{\Theta }$; i.e., for all $\mathbf{\theta }^{i}\in \mathbf{%
\Theta }$, $\mathbf{b}(\mathbf{\theta }^{i}):=\arg \min_{\beta \in \mathbf{B}%
}Q_{\infty }(\mathbf{\theta }^{i}\mathbf{;\beta })$ and $\mathbf{\beta }%
^{0}:=\mathbf{b}(\mathbf{\theta }^{0}).$
\end{description}

\begin{description}
\item[{[G3]}] $\mathbf{b}\left( \mathbf{\cdot }\right) $ is one-to-one in $%
\mathbf{\theta }^{i}$; i.e., $\mathbf{\beta }=\mathbf{b}(\mathbf{\theta }%
^{i})$ has a unique solution for all $\mathbf{\theta }^{i}\in \mathbf{\Theta 
}$.
\end{description}

\noindent If [G1]-[G3] and [P] are satisfied Pr$_{\varepsilon }^{Q}(\mathbf{%
\theta }\notin \aleph _{\delta }(\mathbf{\theta }^{0})|\mathbf{\eta (y))}%
\rightarrow _{p}0$ as $T\rightarrow \infty $ and $\varepsilon \rightarrow 0.$
\end{theorem}

\begin{description}
\item[Remark 10] The above result states that, so long as $Q(\mathbf{.};%
\mathbf{\beta })$ satisfies standard properties ([G1], [G2]), and if the
so-called binding function $\mathbf{b}(\mathbf{\cdot })$ is one-to-one, an
ABC algorithm that uses as summary statistics the minimizers of $Q(.;\mathbf{%
\beta })$ will yield a posterior that is degenerate at $\mathbf{\theta }^{0}$%
. For a specific objective function, conditions [G1] and [G2] are generally
satisfied under more primitive conditions; see, for example, Jennrich (1969)
in the setting where $Q(.;\mathbf{\beta })$\textbf{\ }is the nonlinear least
squares criterion, and Newey and McFadden (1994) in the case where $Q(.;%
\mathbf{\beta })$ is a minimum distance criterion. While the result of
Theorem \ref{thm2} is intuitive it is nonetheless important as it
illustrates that we are not confined to using simple summary statistics of
the data or the log-likelihood function $L\left( \mathbf{y};\mathbf{\beta }%
\right) $ of the auxiliary model within ABC. Instead, any criterion function
satisfying [G1]-[G3] can be used to generate valid summary statistics for
use in ABC.

\item[Remark 11] An alternative to Algorithm 5\textit{\ }is to replace the
summary statistics $\mathbf{\eta (y)}=\widehat{\mathbf{\beta }}(\mathbf{y})$%
, $\mathbf{\eta }(\mathbf{z}^{i})=\widehat{\mathbf{\beta }}(\mathbf{z}^{i})$
in Step (3) with a distance based on $(\partial /\partial \beta )Q(\mathbf{z}%
^{i};\mathbf{\widehat{\beta }(y)})$; e.g., 
\begin{equation}
d\left\{ \mathbf{\eta (y),\eta (z}^{i}\mathbf{)}\right\} =\sqrt{\left\{
(\partial /\partial \mathbf{\beta })Q(\mathbf{z}^{i};\mathbf{\widehat{\beta }%
(y)})\right\} ^{^{\prime }}\widehat{\Omega }(\mathbf{y})\left\{ (\partial
/\partial \mathbf{\beta })Q(\mathbf{z}^{i};\mathbf{\widehat{\beta }(y)}%
)\right\} \text{,}}  \label{score_1}
\end{equation}%
for some positive definite weighting matrix\textbf{\ }$\widehat{\Omega }(%
\mathbf{y}).$ Such an algorithm would be quite useful in situations where $%
(\partial /\partial \beta )Q(\mathbf{z}^{i};\mathbf{\widehat{\beta }(y)})$
is known in closed form and would (in all cases)\textbf{\ }lead to an ABC
algorithm that is several orders of magnitude faster than one based on
computing $\mathbf{\widehat{\beta }}(\mathbf{z}^{i})$ at every value $%
\mathbf{\theta }^{i}$. Under conditions similar to those in Theorem \ref%
{thm2}, a consistency result will hold for the posterior obtained from an
ABC algorithm that uses the distance measure in \eqref{score_1}. We omit
this proof for brevity.
\end{description}

\subsection{The Role of the Auxiliary Model}

Intimately tied to the idea of choosing a suitable criterion function is the
choice of the auxiliary model from which the criterion function is computed.
If the chosen auxiliary model is a poor representation of the observed data
it is likely that no criterion function, likelihood or otherwise, will
produce adequate summary statistics upon which to base our ABC algorithm. In
this way using summary statistics from an auxiliary model inside of ABC is
not a panacea.

ABC algorithms based\textbf{\ }on an auxiliary model and with summary
statistics derived from a criterion function $Q(.;\mathbf{\beta })$ can fail
for precisely the same reason ABC based on arbitrary summary statistics can
fail, namely, failure of [G3] (respectively [S2(2)]). Satisfaction of [G3]
is affected by both\textbf{\ }the choice of the auxiliary model and the
subsequent criterion function used to obtain $\mathbf{\eta (y)=\mathbf{%
\widehat{\beta }(}y\mathbf{)}}$. Since the choice of auxiliary model and
criterion $Q(.;\mathbf{\beta })$ are user and example specific, attempting
to give hard and fast guidelines for how one should choose either is a
research topic in its own right. Rather, we\ simply advocate that validation
of [G1]-[G3] should at least be attempted for any specified combination (of
model and criterion function) before implementing an ABC algorithm. In the
following example we provide support for this statement by illustrating a
case in which consistency is not yielded via what seems to be a sensible ABC
specification: namely the use of an AR(2) auxiliary model along with an OLS
criterion function to produce inference about the true parameters of a MA(2)
model.

\begin{example}
Consider again the MA(2) model from Example 2. Instead of a summary
statistic based ABC approach, consider implementing ABC using summary
statistics generated via the OLS criterion function for the AR(2) auxiliary
model: $y_{t}=\beta _{1}y_{t-1}+\beta _{2}y_{t-2}+\nu _{t}$, with\textbf{\ }$%
\nu _{t}\sim (0,1).$\textbf{\ }Using 
\begin{equation}
Q(\mathbf{y};\mathbf{\beta })=\frac{1}{T}\sum_{t=3}^{T}(y_{t}-\beta
_{1}y_{t-1}-\beta _{2}y_{t-2})^{2},  \label{ols}
\end{equation}%
and OLS estimator $\mathbf{\widehat{\beta }(y)}=(\widehat{\beta }_{1}\mathbf{%
(y)},\widehat{\beta }_{2}\mathbf{(y)})^{\prime }$, the summary statistic $%
\mathbf{\eta }(\mathbf{y})=(\widehat{\beta }_{1}(\mathbf{y}),\widehat{\beta }%
_{2}(\mathbf{y}))^{\prime }$, which has a simple closed form, can be used to
build a computationally simple ABC algorithm.

Given the particular structure of $Q(\cdot ;\mathbf{\beta })$ in (\ref{ols})
and under conditions \eqref{const1} for $\mathbf{\theta }$, [G1] and [G2]
are satisfied. Therefore, all that remains is to verify [G3].
Differentiating the limit criterion $Q_{\infty }(\mathbf{\theta }^{i}\mathbf{%
;\beta })$ with respect to $\mathbf{\beta }=(\beta _{1},\beta _{2})^{\prime
} $ yields the following equations%
\begin{equation}
\begin{array}{c}
E(y_{t-1}(z_{t}^{i}-\beta _{1}z_{t-1}^{i}-\beta _{2}z_{t-2}^{i}))=0, \\ 
E(y_{t-2}(z_{t}^{i}-\beta _{1}z_{t-1}^{i}-\beta _{2}z_{t-2}^{i}))=0.%
\end{array}
\label{zma1}
\end{equation}%
Defining the autocovariances based on $\mathbf{\theta }^{i}$ as $\gamma
_{0}^{i}=(1+(\theta _{1}^{i})^{2}+(\theta _{2}^{i})^{2}),\gamma
_{1}^{i}=(\theta _{1}^{i}+\theta _{1}^{i}\theta _{2}^{i})$ and $\gamma
_{2}^{i}=\theta _{2}^{i}$, we can re-write \eqref{zma1} as 
\begin{equation}
\begin{array}{c}
\gamma _{1}^{i}-\beta _{1}\gamma _{0}^{i}-\beta _{2}\gamma _{1}^{i}=0, \\ 
\gamma _{2}^{i}-\beta _{1}\gamma _{1}^{i}-\beta _{2}\gamma _{0}^{i}=0.%
\end{array}
\label{zma2}
\end{equation}%
Solving for $\beta _{1},\beta _{2}$ in \eqref{zma2} yields the following: 
\begin{equation*}
\beta _{1}(\mathbf{\theta }^{i}):=\left[ \gamma _{1}^{i}-\left( \frac{\gamma
_{1}^{i}\gamma _{2}^{i}}{\gamma _{0}^{i}}\right) \right] /\left[ \gamma
_{0}^{i}-\left( \frac{(\gamma _{1}^{i})^{2}}{\gamma _{0}^{i}}\right) \right]
,\beta _{2}(\mathbf{\theta }^{i}):=\frac{\gamma _{2}^{i}}{\gamma _{0}^{i}}-%
\frac{\gamma _{1}^{i}}{\gamma _{0}^{i}}\beta _{1}(\mathbf{\theta }^{i}).
\end{equation*}%
Interestingly, and\ as an illustration of the point made in Section \ref%
{prelim}, the binding function $\mathbf{b}(\mathbf{\theta }^{i})=(\beta _{1}(%
\mathbf{\theta }^{i}),\beta _{2}(\mathbf{\theta }^{i}))^{\prime }$ does not
admit a unique solution to $0=\mathbf{\mathbf{b(\mathbf{\theta }^{0})}%
-b(\theta }^{i}\mathbf{)},$\textbf{\ }for all\textbf{\ }$\mathbf{\theta }%
^{0},\mathbf{\theta }^{i}$\textbf{$\in \mathbf{\Theta }$.} For instance,
simple numerical calculations reveal that if $\mathbf{\theta }^{0}=(\theta
_{1}^{0},\theta _{2}^{0})^{\prime }=\left( .6,.2\right) ^{\prime }$ the
equation $0=\mathbf{\mathbf{b(\mathbf{\theta }^{0})}-b(\theta }^{i}\mathbf{)}
$ has a unique solution satisfying the conditions of \eqref{const1}, namely $%
\mathbf{\theta }^{i}\mathbf{=\theta }^{0}$ (a second solution, $\mathbf{%
\theta }^{i}=(3,5)^{\prime }$, exists but does not satisfy the parameter
restrictions \eqref{const1}). However, if\textbf{\ } $\mathbf{\theta }%
^{0}=\left( .5,.5\right) ^{\prime },$ the equation $0=\mathbf{\mathbf{b(%
\mathbf{\theta }^{0})}-b(\theta }^{i}\mathbf{)}$ has two solutions
satisfying the conditions of \eqref{const1}, $\mathbf{\theta }^{i}\mathbf{=}%
\left( .5,.5\right) ^{\prime }$ and $\mathbf{\theta }^{i}=(1,2)$! Therefore, 
$\mathbf{b}(\mathbf{\theta }^{i})=(\beta _{1}(\mathbf{\theta }^{i}),\beta
_{2}(\mathbf{\theta }^{i}))^{\prime }$ is not a one-to-one function and
hence will not yield consistent inference in general.
\end{example}

\section{Consistency of ABC in Ordinary Differential Equations Models\label%
{ode_section}}

In this section we investigate the ability of ABC to yield Bayesian
consistent inference for parameters governing a system of ordinary
differential equations (ODEs). As will be demonstrated, this particular type
of application, which has been given some attention in the ABC literature {%
(see, for example, Toni \textit{et al.}, 2009, Sun \textit{et al.}, 2014,
Prangle, 2015),} highlights certain important issues related to Bayesian
consistency of ABC-based posterior estimates. In particular, by checking the
conditions of Theorem \ref{thm2} in a simple deterministic system, we
demonstrate that ABC can yield inconsistent inference in such settings,
highlighting the importance of these conditions for verifying the validity
of ABC-based inference. While we specifically focus on a simple
deterministic system, these findings can easily be generalized to other ODEs.

Specifically, we give our attention to the Lotka-Volterra (LV) model, which
describes the interaction between a species $x_{1}$, referred to as the prey
species, and a species $x_{2}$, referred to as the predator species. For ${%
\mathbf{\theta }}=(\theta _{1},\theta _{2})^{\prime }$ unknown, we consider
the deterministic LV model defined through the system of ODEs: 
\begin{eqnarray}
\frac{dx_{1}}{dt} &=&\theta _{1}x_{1}-x_{1}x_{2},  \label{ODE} \\
\frac{dx_{2}}{dt} &=&\theta _{2}x_{1}x_{2}-x_{2}.  \notag
\end{eqnarray}%
For any point $t_{i}$ in the interval $[0,T],$ the vector ${\mathbf{x}}%
(t_{i})=(x_{1}(t_{i}),x_{2}(t_{i}))^{\prime }$ is the solution to the above
ODEs, with initial value $t_{0}.$ Typically, it is assumed that we do not
observe ${\mathbf{x}}(t_{i})$; rather, we observe a quantity corresponding
to ${\mathbf{x}}(t_{i})$ that is measured with error that is both additive
and independent over observational points, see, for example, Beck and Arnold
(1977). Following this usual practice then, we specify a measurement
equation of the form 
\begin{equation}
{\mathbf{y}}(t_{i})={\mathbf{x}}(t_{i})+{\mathbf{\nu }}(t_{i}),
\label{ODEerror}
\end{equation}%
where \textbf{\ }${\mathbf{\nu }}(t_{i})\sim i.i.d.(0,\Sigma _{v})$ and $%
\Sigma _{v}$ is diagonal.\footnote{%
It could be assumed that the evolution of $x_{1}$ and $x_{2}$ is stochastic
rather than deterministic; however the adoption (or not) of this assumption
is not germane to our discussion and we thus retain the ODE structure for
the states.}

Assume we have an observed sample of size $R_{T}$ from \eqref{ODEerror},
with corresponding design points $t_{1},...,t_{R_{T}},$ fixed or random. Our
goal is to estimate the posterior density of ${\mathbf{\theta }}$ using the
observed sample $\left\{ {\mathbf{y}}(t_{i})\right\} _{i=1}^{R_{T}}\,,$ and
prior density $p({\mathbf{\theta }}).$ Toni \textit{et al.} (2009) propose
to estimate these posteriors via ABC using the squared distance between the
observed and simulated samples. {Specifically}, for $\left\{ {\mathbf{y}}%
(t_{i})\right\} _{i=1}^{R_{T}}$ the observed sample and $\left\{ {\mathbf{z}}%
(t_{i};{\mathbf{\theta }})\right\} _{i=1}^{R_{T}}$ the simulated sample,
obtained by solving equation (\ref{ODE}) at ${\mathbf{\theta }}=(\theta
_{1},\theta _{2})^{\prime }$, ABC {is} based on the distance 
\begin{equation}
\rho \left\{ \mathbf{y,z(\theta )}\right\} =\frac{1}{R_{T}}%
\sum_{j=1}^{2}\sum_{i=1}^{R_{T}}(y_{j}(t_{i})-z_{j}(t_{i};\mathbf{\theta }%
)))^{2}.  \label{odeD1}
\end{equation}%
That is, draws of $\mathbf{\theta }$ are retained according to the proximity
of the stochastic quantity $y_{j}(t_{i})$ to the deterministic quantity $%
z_{j}(t_{i};\mathbf{\theta }).$

It is critical to note, however, that choosing values of $\mathbf{\theta }%
^{i}$ such that $\rho \left\{ \mathbf{y,z(\theta }^{i}\mathbf{)}\right\}
\leq \varepsilon $ will not yield an ABC-based posterior that is Bayesian
consistent. This can be seen by noting that as $R_{T}\rightarrow \infty ,$
even if we select $\mathbf{\theta }^{i}\mathbf{=\theta }^{0}$, and so $%
x_{j}(t_{i})=z_{j}(t_{i};\mathbf{\theta }^{0})$ for all $t_{i}$ and $j=1,2$,
it will be the case that $\lim_{R_{T}\rightarrow \infty }\rho \left\{ 
\mathbf{y,z(\theta }^{0}\mathbf{)}\right\} \overset{P}{\rightarrow }E(\nu
_{1}^{2}(t_{i})+\nu _{2}^{2}(t_{i}))>\varepsilon $, for $\varepsilon $
arbitrarily small. Therefore, there exists no value of $\mathbf{\theta }%
^{i}\in \mathbf{\Theta }$ for which $\rho \left\{ \mathbf{y,z(\theta }^{i}%
\mathbf{)}\right\} \leq \varepsilon $ as $R_{T}\rightarrow \infty $ and $%
\varepsilon \rightarrow 0,$ and so the ABC-based posterior defined by the
distance\textbf{\ }$\rho \left\{ \mathbf{y,z(\theta )}\right\} $\textbf{\ }%
can not be Bayesian consistent.

However, an alternative to the \textquotedblleft distance\textquotedblright\ 
$\rho \left\{ {\mathbf{y,z(\theta )}}\right\} $ in \eqref {odeD1} is a
metric based on statistics obtained from minimizing an objective function
representing the data in equation (\ref{ODEerror}). A common means of
obtaining (frequentist) point estimates for parameters defined by ODEs is
nonlinear least squares (NLS), whereby the squared distance between the
observed and simulated solutions is minimized (see\textbf{\ }Beck and
Anrnold, 1977, for a discussion). This then motivates us to consider the
consistency properties of an ABC method that mimics the spirit of NLS. As
such, we consider as summary statistics for use in ABC, the parameters that
minimize the ordinary least squares (OLS) criterion 
\begin{equation*}
Q({\mathbf{y;\beta }})=\frac{1}{R_{T}}\sum_{j=1}^{2}\frac{1}{\beta _{2,j}}%
\sum_{i=1}^{R_{T}}(y_{j}(t_{i})-\beta _{1,j})^{2},
\end{equation*}%
with respect to $\mathbf{\beta }=(\mathbf{\beta }_{1}^{\prime },\mathbf{%
\beta }_{2}^{\prime })^{\prime }$, ${\mathbf{\beta }}_{1}=({\beta }_{1,1},{%
\beta }_{1,2})^{\prime }$ and ${\mathbf{\beta }}_{2}=({\beta }_{2,1},{\beta }%
_{2,2})^{\prime },$ which defines $\widehat{\beta }_{1,j}=\frac{1}{R_{T}}%
\sum_{i=1}^{R_{T}}y_{j}(t_{i})$ as the sample mean and $\widehat{\beta }%
_{2,j}=\frac{1}{R_{T}}\sum_{i=1}^{R_{T}}(y_{j}(t_{i})-\widehat{\beta }%
_{1,j})^{2}$ as the sample variance. ABC can then be conducted\textbf{\ }%
using $\widehat{{\mathbf{\beta }}}(\mathbf{y})=(\widehat{{\mathbf{\beta }}}%
_{1}^{\prime }(\mathbf{y}),\widehat{{\mathbf{\beta }}}_{2}^{\prime }(\mathbf{%
y}))^{\prime }$ and its simulated counterpart $\widehat{{\mathbf{\beta }}}(%
\mathbf{z}^{i})=(\widehat{{\mathbf{\beta }}}_{1}^{\prime }(\mathbf{z}^{i}),%
\widehat{{\mathbf{\beta }}}_{2}^{\prime }(\mathbf{z}^{i}))^{\prime }\,,$
with a distance of the form specified in (\ref{general_distance}) adopted.
Further alternatives can be defined by basing ABC on matching $\widehat{{%
\mathbf{\beta }}}_{1}(\mathbf{y})$ alone (respectively, $\widehat{{\mathbf{%
\beta }}}_{2}(\mathbf{y})$) with its simulated counterpart $\widehat{{%
\mathbf{\beta }}}_{1}(\mathbf{z}^{i})$ (respectively, $\widehat{{\mathbf{%
\beta }}}_{2}(\mathbf{z}^{i})$), with the use of $\widehat{{\mathbf{\beta }}}%
_{1}(\mathbf{\cdot })$ alone as the matching statistic being closest in
spirit to NLS.

Sufficient conditions guaranteeing that ABC will yield consistent inference
are given in Theorem \ref{thm2} and must be verified, for each version of ${%
\mathbf{\eta (y)}},{\mathbf{\eta (z}^{i}})$ obtained from $Q({\mathbf{\cdot
;\beta }}).$ {Whilst formal verification} of the identification condition in
this case is complicated by the fact that ${\mathbf{x}}(t_{i})$ has no
closed form, some analytical insights are attainable, by noting the
following. Define $\bar{x}_{j}=\lim_{R_{T}\rightarrow \infty }\frac{1}{R_{T}}%
\sum_{i=1}^{R_{T}}x_{j}(t_{i}),$ $\bar{x}_{j}^{2}=\lim_{R_{T}\rightarrow
\infty }\frac{1}{R_{T}}\sum_{i=1}^{R_{T}}x_{j}^{2}(t_{i}),$ and
corresponding simulated counterparts $\bar{z}_{j}({\mathbf{\theta }}%
)=\lim_{R_{T}\rightarrow \infty }\frac{1}{R_{T}}%
\sum_{i=1}^{R_{T}}z_{j}(t_{i};{\mathbf{\theta }}),$ $\bar{z}_{j}^{2}({%
\mathbf{\theta }})=\lim_{R_{T}\rightarrow \infty }\frac{1}{R_{T}}%
\sum_{i=1}^{R_{T}}z_{j}^{2}(t_{i};{\mathbf{\theta }}),$ for $j=1,2$.
Assuming these quantities exist, it can be shown that%
\begin{equation*}
\lim_{R_{T}\rightarrow \infty }\widehat{\beta }_{2,j}(\mathbf{y})%
\xrightarrow {P}\bar{x}_{j}^{2}+\sigma _{j}^{2}-(\bar{x}_{j})^{2},
\end{equation*}%
where $\sigma _{j}^{2}$ is the $(j,j)$ element of $\Sigma _{v}$. However, it
is also the case that\textbf{\ }%
\begin{equation*}
\lim_{R_{T}\rightarrow \infty }\widehat{\beta }_{2,j}({\mathbf{z({\theta })}}%
)\xrightarrow {P}\bar{z}_{j}^{2}({\mathbf{\theta }})-\left( \bar{z}_{j}({%
\mathbf{\theta }})\right) ^{2}.
\end{equation*}%
Hence, as $R_{T}\rightarrow \infty $%
\begin{equation*}
\left\Vert \widehat{{\mathbf{\beta }}}_{2}(\mathbf{y})-\widehat{{\mathbf{%
\beta }}}_{2}(\mathbf{z(}{\mathbf{\theta }}^{0}))\right\Vert \neq o_{P}(1)
\end{equation*}%
and so there is no hope that ABC based on ${\mathbf{\eta (y)}}=\widehat{{%
\mathbf{\beta }}}_{2}(\mathbf{y})$ will yield consistent inference. That is,
and reverting to the general notation of the previous section, $Q({\mathbf{y}%
};{\mathbf{\beta }})$ and $Q({\mathbf{z}}(\mathbf{\theta }^{0});{\mathbf{%
\beta }})$ do not have corresponding limit $Q_{\infty }(\mathbf{\theta }^{0};%
{\mathbf{\beta }})$, which violates Assumption [G1] of Theorem \ref{thm2}.
This same point pertains to the case in which the augmented statistic $%
\widehat{{\mathbf{\beta }}}(\mathbf{y})$ is used.

The critical insight from the above illustration, as it pertains to ABC, is
that a mismatch between the assumed processes for the observed and simulated
data, with the latter failing to replicate the stochastic nature of the
former, can create a fundamental disconnect between the matching statistics
formed from the two sets of data, so that they will never coincide, no
matter the proximity of the drawn parameter vector to the truth. We now
contrast this with an alternative approach in which we deliberately draw
simulated data according to the measurement equation 
\begin{equation}
{\mathbf{z}}(t_{i};{\mathbf{\theta }})={\mathbf{w}}(t_{i};{\mathbf{\theta }}%
)+{\mathbf{\tilde{\nu}}}(t_{i}),  \label{simODE}
\end{equation}%
where ${\mathbf{w}}(t_{i};{\mathbf{\theta }})$ is the numerical solution of
the ODE at parameter value ${\mathbf{\theta }}$ and ${\mathbf{\tilde{\nu}}}%
(t_{i})$ is a random error drawn from the same distribution as ${\mathbf{\nu 
}}(t_{i}).$ {In this case, it} is easy to verify that the simulated
statistics $\widehat{{\mathbf{\beta }}}_{2}(\mathbf{z}^{i})\ $and $\widehat{{%
\mathbf{\beta }}}(\mathbf{z}^{i})$ depend on {the measurement error variance 
}$\Sigma _{v}${\ }in the same manner as the observed data, with Assumption
[G1] of Theorem \ref{thm2} no longer violated as a consequence, and so,%
\textbf{\ }as $R_{T}\rightarrow \infty $\textbf{\ } 
\begin{equation*}
\left\Vert \widehat{{\mathbf{\beta }}}_{2}(\mathbf{y})-\widehat{{\mathbf{%
\beta }}}_{2}(\mathbf{z(}{\mathbf{\theta }}^{0}))\right\Vert =o_{P}(1),{%
\text{ and }}\left\Vert \widehat{{\mathbf{\beta }}}(\mathbf{y})-\widehat{{%
\mathbf{\beta }}}(\mathbf{z(}{\mathbf{\theta }}^{0}))\right\Vert =o_{P}(1).
\end{equation*}%
Once again, since no closed form solution exists for the state process,
establishing the identification condition analytically, as in the previous
examples, is not feasible. However, numerical exploration indicates the
existence of consistency for matching statistics $\widehat{{\mathbf{\beta }}}%
_{1}(y),$ $\widehat{{\mathbf{\beta }}}_{2}(y)$ and $\widehat{{\mathbf{\beta }%
}}(y)$\ when data is simulated according to\textbf{\ }\eqref {simODE}.%
\footnote{%
Numerical results illustrating consistency are available from the authors
upon request.}

It has generally been recognized that consistent inference for point
estimates of parameters in ODEs is due to the additive nature of the
measurement error, combined with the fact that the measurement error has
mean zero, known variance, and is independent of the data, as well as the
satisfaction of identification conditions guaranteeing the existence of a
unique minimum at\textbf{\ }$\mathbf{\theta }^{0}$; see, for example, Beck
and Arnold (1977). However, when conducting inference for ODEs via ABC, we
see that\textbf{\ }in addition to these conditions (or variants thereof),
care must be taken to ensure that data is simulated in such a way that it
matches the observed data. It is the price we pay for conducting complete
inference using a simulation-based procedure.

\section{Discussion \label{disc}}

Consistency is one of the most fundamental properties with which to gauge
the output of a statistical inference procedure. With our focus on Bayesian
consistency, we demonstrate that in the limit (as both $T\rightarrow \infty $
and $\varepsilon \rightarrow 0$) the ABC posterior estimate will be
degenerate at the true parameter (vector) if (and only if!) the summary
statistics upon which ABC is based are appropriately chosen. Conditions
guaranteeing Bayesian consistency of ABC posterior estimates for a wide
range of summary statistics, with and without respect to an auxiliary model,
with the former defined with respect to an arbitrary criterion function, are
developed and several examples featured in the literature are used to
illustrate these conditions. The results are less heartening than expected
and demonstrate that consistent inference in ABC is in no way guaranteed. In
general, we find that ABC will only yield consistent inference when a
judicious choice of summary statistics has been employed, subsequently
calling into question a large collection of ABC results based on arbitrary
summary statistics, as well as those generated from well-specified auxiliary
models. In addition, our results highlight the need both to specify a proper
distance measure and to ensure an exact match between the process assumed to
have generated the observed data and that used to produce simulated samples,
in order to have any hope of yielding consistent inference.

To determine if ABC\ will be Bayesian consistent in practice, we develop a
useful and computationally simple diagnostic procedure that can be applied
to any given data set and any choice of summary statistics. This procedure
constitutes an important first step in determining, in any practical
situation, whether ABC will yield consistent inference. Formalization of
this diagnostic procedure, as well as work detailing its theoretical
properties, is a topic of ongoing research by the authors.

Before closing, we re-emphasize the fact that the results presented herein,
while cast within the framework of the ABC accept/reject algorithm, apply to
the more sophisticated variants of the ABC method. In particular, the
results are applicable to ABC algorithms that generate summary statistics
through various simulation-based approximations, as well as algorithms that
utilize more efficient methods of post-sampling density estimation. Given
this fact, the results discussed herein can be used to form the basic
foundation for determining Bayesian consistency for all summary
statistic-based ABC algorithms. Further, the key issue that we have
emphasized throughout, namely the need to verify the relevant conditions for
Bayesian consistency, including the required one-to-one property of the
(implied) binding function, is just as pertinent, of course, to related
frequentist simulation-based inference methods. In particular, the
development of a formal and rigorous method for confirming the one-to-one
property of the binding function is as critical to the establishment of the
(asymptotic) validity of all other such methods as it is to ABC.

\section{Appendix: Proofs}

\begin{proof}[Proof of Theorem 1]
The proof is broken into three parts: first, we show that the only value $%
\mathbf{\theta }^{i}$ that will be selected for all $\varepsilon \geq 0$ as $%
T\rightarrow \infty $ is $\mathbf{\theta }^{i}=\mathbf{\theta }^{0}$;
second, we demonstrate that for any $\varepsilon >0$ there exists some $%
N(\varepsilon )$ such that if $N>N(\varepsilon )$ the posterior density $%
p_{\varepsilon }(\mathbf{\theta }|\mathbf{\eta }(\mathbf{y}))$ has a
well-defined probability limit; lastly, we use these two pieces to
demonstrate that for any $\aleph _{\delta }(\mathbf{\theta }^{0})$ and $%
\Lambda _{\delta }:=\Theta /\aleph _{\delta }(\mathbf{\theta }^{0})$, the
posterior probability $\text{Pr}_{\varepsilon }(\mathbf{\theta }\in \Lambda
_{\delta }|\mathbf{\eta }(\mathbf{y}))\xrightarrow{P}0$.

\bigskip \noindent {\textbf{\underline{Part 1:}}}\bigskip \noindent

By the triangle inequality 
\begin{equation}
d\{\mathbf{\eta }(\mathbf{y}),\mathbf{\eta }(\mathbf{z}^{i})\}\leq d\{%
\mathbf{\eta }(\mathbf{y}),\mathbf{b}(\mathbf{\theta }^{i})\}+d\{\mathbf{b}(%
\mathbf{\theta }^{i}),\mathbf{\eta }(\mathbf{z}^{i})\}.  \label{star_0}
\end{equation}%
Applying the triangle inequality again to the first term on the
right-hand-side of \eqref{star_0} yields 
\begin{equation*}
d\{\mathbf{\eta }(\mathbf{y}),\mathbf{\eta }(\mathbf{z}^{i})\}\leq d\{%
\mathbf{b}(\mathbf{\theta }^{0}),\mathbf{b}(\mathbf{\theta }^{i})\}+d\{%
\mathbf{\eta }(\mathbf{y}),\mathbf{b}(\mathbf{\theta }^{0})\}+d\{\mathbf{b}(%
\mathbf{\theta }^{i}),\mathbf{\eta }(\mathbf{z}^{i})\}.
\end{equation*}%
By [S1], $\mathbf{\eta }(\mathbf{y})\overset{P}{\rightarrow }\mathbf{b}(%
\mathbf{\theta }^{0})$ and so $d\{\mathbf{\eta }(\mathbf{y}),\mathbf{b}(%
\mathbf{\theta }^{0})\}=o_{P}(1)$. In addition, 
\begin{equation*}
d\{\mathbf{\eta }(\mathbf{z}^{i}),\mathbf{b}(\mathbf{\theta }^{i})\}\leq
\sup_{\mathbf{\theta \in \Theta }}d\{\mathbf{\eta }(\mathbf{z(\theta )}),%
\mathbf{b}(\mathbf{\theta })\},
\end{equation*}%
and by [S2(1)] $\sup_{\mathbf{\theta \in \Theta }}d\{\mathbf{\eta }(\mathbf{%
z(\theta )}),\mathbf{b}(\mathbf{\theta })\}=o_{P}(1)$. Combining these facts
yields 
\begin{equation}
d\{\mathbf{\eta }(\mathbf{y}),\mathbf{\eta }(\mathbf{z}^{i})\}\leq d\{%
\mathbf{b}(\mathbf{\theta }^{0}),\mathbf{b}(\mathbf{\theta }^{i})\}+o_{P}(1).
\label{uselats}
\end{equation}%
For fixed $\varepsilon >0,$ as $T\rightarrow \infty $ a value $\mathbf{%
\theta }^{i}$ will be selected if 
\begin{equation*}
d\{\mathbf{b}(\mathbf{\theta }^{0}),\mathbf{b}(\mathbf{\theta }%
^{i})\}+o_{P}(1)\,\,\leq \varepsilon .
\end{equation*}%
By [S2(2)] the only value of $\mathbf{\theta }^{i}\in \mathbf{\Theta }$ for
which $\mathbf{b}(\mathbf{\theta }^{i})=\mathbf{b}(\mathbf{\theta }^{0})$ is 
$\mathbf{\theta }^{i}=\mathbf{\theta }^{0}$. Therefore, as $T\rightarrow
\infty $, the only value of $\mathbf{\theta }^{i}$ satisfying $d\{\mathbf{b}(%
\mathbf{\theta }^{0}),\mathbf{b}(\mathbf{\theta }^{i})\}\leq \varepsilon $
for any $\varepsilon \geq 0$ is $\mathbf{\theta }^{i}=\mathbf{\theta }^{0}$.

\bigskip \noindent {\textbf{\underline{Part 2:} }}

\bigskip \noindent Part 1 suggests that for small enough $\varepsilon $ the
posterior density $p_{\varepsilon }(\mathbf{\theta }|\mathbf{\eta }(\mathbf{y%
}))$ will be zero for values $\mathbf{\theta }$ $\in \Lambda _{\delta }$ as $%
T\rightarrow \infty $. However, because $p_{\varepsilon }(\mathbf{\theta }|%
\mathbf{\eta }(\mathbf{y}))$ is built from $N$ random draws, for any $%
\varepsilon >0$ we must ensure that $N$ can be chosen large enough so that $%
\text{plim}_{T\rightarrow \infty }p_{\varepsilon }(\mathbf{\theta }|\mathbf{%
\eta }(\mathbf{y}))$ exists for any $\varepsilon >0$.

By compactness of $\mathbf{\Theta }$ and Assumption [P], for any $r>0$ there
exists a finite integer $N(r)$ and points $\{\mathbf{\theta }%
^{i}\}_{i=1}^{N(r)}$, each drawn according to $\mathbf{\theta }^{i}\sim p(%
\mathbf{\theta })$, such that 
\begin{equation*}
\mathbf{\Theta }=\bigcup_{i=1}^{N(r)}\aleph _{r}(\mathbf{\theta }^{i})\text{
w.p.1.}
\end{equation*}%
By continuity of $\mathbf{b}(\mathbf{\cdot })$, there exists an $%
r(\varepsilon )>0$ such that, $\Vert \mathbf{\theta }-\mathbf{\theta }%
^{0}\Vert <r(\varepsilon )$ implies $d\{\mathbf{b}(\mathbf{\theta }^{0}),%
\mathbf{b}(\mathbf{\theta })\}<\varepsilon $ for any $\varepsilon >0$.
Combining the two ideas we see that for any $\varepsilon >0$, we can cover $%
\mathbf{\Theta }$ with $N(r(\varepsilon ))$ balls w.p.1. Now, note that by
Assumption [P] and the above argument, for any $\varepsilon >0$, we can find
a radius $r(\varepsilon )$ such that $\mathbf{\theta }^{0}\in \aleph
_{r(\varepsilon )}(\mathbf{\theta }^{i})$ for some $\mathbf{\theta }^{i}$,
and $d\{\mathbf{b}(\mathbf{\theta }^{0}),\mathbf{b}(\mathbf{\theta }%
^{i})\}<\varepsilon \text{{}}\text{ w.p.1}$ as a consequence. In addition,
by Assumption [S1], [S2(1)] and Part 1 of the proof, 
\begin{equation*}
d\{\mathbf{\eta (y)},\mathbf{\eta (z(\theta }^{i}))\}=d\{\mathbf{b(\theta }%
^{0}),\mathbf{b(\theta }^{i})\}+o_{P}(1)
\end{equation*}%
for any $\mathbf{\theta }^{i}\in \mathbf{\Theta }$ and so 
\begin{equation*}
d\{\mathbf{\eta (y)},\mathbf{\eta (z(\theta }^{i}))\}<\varepsilon +o_{P}(1)
\end{equation*}%
for $\mathbf{\theta }^{i}$ such that $\mathbf{\theta }^{0}\in \aleph
_{r(\varepsilon )}(\mathbf{\theta }^{i})$. Using this fact, we have that 
\begin{eqnarray*}
\mathbb{I}_{\varepsilon }[\mathbf{z}(\mathbf{\theta }^{i})]=\mathbb{I}[d\{%
\mathbf{\eta (y),\eta (z(\theta }^{i}))\}\leq \varepsilon ] &=&\mathbb{I}[d\{%
\mathbf{b}(\mathbf{\theta }^{0}),\mathbf{b}(\mathbf{\theta }^{i})\}\leq
\varepsilon ]+o_{P}(1), \\
&=&1+o_{P}(1).
\end{eqnarray*}%
From here we see that for $T$ arbitrarily large and any $\varepsilon >0$,
there exists some $N(\varepsilon ):=N(r(\varepsilon ))$ such that for $%
N>N(\varepsilon )$ 
\begin{equation}
p_{\varepsilon }(\mathbf{\theta }|\mathbf{\eta }(\mathbf{y}))=\int_{\mathbf{z%
}}\frac{p(\mathbf{\theta })p(\mathbf{z}|\mathbf{\theta })\mathbb{I}_{{%
\varepsilon }}[\mathbf{z}(\mathbf{\theta })]}{\left\{ \int_{\mathbf{\Theta }%
}\int_{\mathbf{z}}p(\mathbf{\theta })p(\mathbf{z}|\mathbf{\theta })\mathbb{I}%
_{A_{\varepsilon }}[\mathbf{z}(\mathbf{\theta })]d\mathbf{z}d\mathbf{\theta }%
\right\} }d\mathbf{z}  \label{pedef}
\end{equation}%
exists.

\bigskip

\noindent{\textbf{\underline{Part 3:}}}

\bigskip \noindent We now use Parts 1 and 2 to show that $\text{Pr}%
_{\varepsilon }(\mathbf{\theta }\in \Lambda _{\delta }|\mathbf{\eta (y)})%
\xrightarrow{P}0$, where $\Lambda _{\delta }:=\mathbf{\Theta }/\aleph
_{\delta }(\mathbf{\theta }^{0})$. By Markov's inequality 
\begin{equation}
\lim_{T\rightarrow \infty }\text{Pr}\bigg\{\text{Pr}_{\varepsilon }(\mathbf{%
\theta }\in \Lambda _{\delta }|\mathbf{\eta }(\mathbf{y}))>\xi \bigg\}\leq
\lim_{T\rightarrow \infty }\text{E}\bigg\{\text{Pr}_{\varepsilon }(\mathbf{%
\theta }\in \Lambda _{\delta }|\mathbf{\eta }(\mathbf{y}))\bigg\}/\xi ,
\label{mineq}
\end{equation}%
for all $\xi >0$, and the result follows if the left-hand side of %
\eqref{mineq} is zero. By the definition of $p_{\varepsilon }(\mathbf{\theta 
}|\mathbf{\eta }(\mathbf{y}))$, $\text{Pr}_{\varepsilon }(\mathbf{\theta }%
\in A|\mathbf{\eta }(\mathbf{y}))<1$ for any $A\subset \mathbf{\Theta }$, $%
\varepsilon >0$, and $T\geq 1$. By the bounded convergence theorem%
\begin{equation*}
\lim_{T\rightarrow \infty }\text{E}_{{}}\left[ \text{Pr}_{\varepsilon }(%
\mathbf{\theta }\in \Lambda _{\delta }|\mathbf{\eta }(\mathbf{y}))\right] =%
\text{E}_{{}}\left[ \text{plim}_{T\rightarrow \infty }\text{Pr}_{\varepsilon
}(\mathbf{\theta }\in \Lambda _{\delta }|\mathbf{\eta }(\mathbf{y}))\right] .
\end{equation*}

By the definition of $p_{\varepsilon }(\mathbf{\theta }|\eta (\mathbf{y}))$
in \eqref{pedef}, $\text{Pr}_{\varepsilon }(\mathbf{\theta }\in \Lambda
_{\delta }|\mathbf{\eta }(\mathbf{y}))=o_{P}(1)$ only if, for some, $%
\varepsilon >0$ 
\begin{equation*}
\sup_{\theta \in \Lambda _{\delta }}\mathbb{I}_{{}}\left[ d\{\mathbf{\eta }(%
\mathbf{y}),\mathbf{\eta }(\mathbf{z}(\mathbf{\theta }))\}\leq \varepsilon %
\right] =o_{P}(1).
\end{equation*}%
For any $\delta >0$, if $\Vert \mathbf{\theta }-\mathbf{\theta }^{0}\Vert
\geq \delta $, by injectivity of $\mathbf{b}(\cdot )$, it follows that $d\{%
\mathbf{b}(\mathbf{\theta }),\mathbf{b}(\mathbf{\theta }^{0})\}\geq
\varepsilon _{\ast }$ for some $\varepsilon _{\ast }>0$. By compactness of $%
\Lambda _{\delta }$ and continuity of $\mathbf{b}(\cdot )$, there exists
some $\mathbf{\theta }_{\ast }$ (not necessarily unique) such that 
\begin{equation}
\mathbf{\theta }_{\ast }=\arg \inf_{\theta \in \Lambda _{\delta }}d\{\mathbf{%
b}(\mathbf{\theta }^{0}),\mathbf{b}(\mathbf{\theta })\},  \label{tstar}
\end{equation}%
and, by injectivity of $\mathbf{b}(\cdot ),$ for some $\varepsilon _{\ast
}>0 $ 
\begin{equation}
d\{\mathbf{b}(\mathbf{\theta }^{0}),\mathbf{b}(\mathbf{\theta }_{\ast
})\}\geq \varepsilon _{\ast }>0.  \label{varstar}
\end{equation}%
Moreover, by Assumptions [S1], [S2(1)], and equation \eqref{uselats} in Part
1, 
\begin{equation}
\sup_{\theta \in \Lambda _{\delta }}|\mathbb{I}_{\varepsilon }[\mathbf{z}(%
\mathbf{\theta })]-\mathbb{I}_{\varepsilon }[\mathbf{b}(\mathbf{\theta }%
)]|=\sup_{\theta \in \Lambda _{\delta }}|\mathbb{I}_{{}}[d\{\mathbf{\eta }(%
\mathbf{y}),\mathbf{\eta }(\mathbf{z}(\mathbf{\theta }))\}\leq \varepsilon ]-%
\mathbb{I}_{{}}[d\{\mathbf{b}(\mathbf{\theta }^{0}),\mathbf{b}(\mathbf{%
\theta })\}\leq \varepsilon ]|=o_{P}(1).  \label{indcon}
\end{equation}%
From equation \eqref{indcon} it follows that 
\begin{equation*}
\mathbb{I}[d\{\mathbf{\eta }(\mathbf{y}),\mathbf{\eta }(\mathbf{z}(\mathbf{%
\theta _{\ast }}))\}\leq \varepsilon ]=\mathbb{I}_{{}}[d\{\mathbf{b}(\mathbf{%
\theta }^{0}),\mathbf{b}(\mathbf{\theta _{\ast }})\}\leq \varepsilon
]+o_{P}(1).
\end{equation*}%
By the definition of $\mathbf{\theta }_{\ast }$ in \eqref{tstar} and
equation \eqref{varstar}, for any $\varepsilon ^{\ast }<\varepsilon _{\ast }$%
, 
\begin{equation*}
\mathbb{I}[d\{\mathbf{b}(\mathbf{\theta }^{0}),\mathbf{b}(\mathbf{\theta
_{\ast }})\}\leq \varepsilon ^{\ast }]=0.
\end{equation*}%
We can then conclude that 
\begin{equation*}
\mathbb{I}[d\{\mathbf{\eta }(\mathbf{y}),\mathbf{\eta }(\mathbf{z}(\mathbf{%
\theta }_{\ast })\}\leq \varepsilon ^{\ast }]=o_{P}(1).
\end{equation*}%
Moreover, from equation \eqref{tstar} 
\begin{equation}
\inf_{\theta \in \Lambda _{\delta }}d\{\mathbf{b}(\mathbf{\theta }^{0}),%
\mathbf{b}(\mathbf{\theta })\}=d\{\mathbf{b}(\mathbf{\theta }^{0}),\mathbf{b}%
(\mathbf{\theta _{\ast }})\}\geq \varepsilon _{\ast }>\varepsilon ^{\ast },
\label{almost}
\end{equation}%
and so it follows from \eqref{indcon} and equation \eqref{almost} 
\begin{equation*}
\sup_{\theta \in \Lambda _{\delta }}\mathbb{I}[d\{\mathbf{\eta }(\mathbf{y}),%
\mathbf{\eta }(\mathbf{z}(\mathbf{\theta }))\}\leq \varepsilon ^{\ast
}]=o_{P}(1).
\end{equation*}%
Therefore, for $\varepsilon \leq \varepsilon ^{\ast }$ and a corresponding $%
N>N(\varepsilon )$ number of simulated draws, which exists by Part 2, $\text{%
Pr}_{\varepsilon }(\mathbf{\theta }\in \Lambda _{\delta }|\mathbf{\eta }(%
\mathbf{y}))=o_{P}(1)$ and the result follows.
\end{proof}

\bigskip

\begin{proof}[Proof of Corollary 1]
We have two cases to consider: one, the vector $\mathbf{b}(\mathbf{\theta }%
^{i})$ is one-to-one in $\mathbf{\theta }^{i}$ and two, only the sub-vector $%
\mathbf{b}_{1}(\mathbf{\theta }^{i})$ is one-to-one in $\mathbf{\theta }^{i}$%
. Clearly,{\ if the first case obtains }then the result follows from Theorem
1 and so we can focus on the latter case.

{For the second case then, by} the triangle inequality 
\begin{equation}
d\{\mathbf{\gamma }(\mathbf{y}),\mathbf{\gamma }(\mathbf{z^{i}})\}\leq d\{%
\mathbf{\gamma }(\mathbf{y}),\mathbf{b}(\mathbf{\theta }^{i})\}+d\{\mathbf{b}%
(\mathbf{\theta }^{i}),\mathbf{\gamma }(\mathbf{z}^{i})\}.  \label{gamma_tri}
\end{equation}%
Using the same arguments as in Theorem \ref{thm1}, equation \eqref{gamma_tri}
can be restated as 
\begin{equation*}
d\{\mathbf{\gamma }(\mathbf{y}),\mathbf{\gamma }(\mathbf{z^{i}})\}\leq d\{%
\mathbf{b}(\mathbf{\theta }^{0}),\mathbf{b}(\mathbf{\theta }^{i})\}+o_{P}(1).
\end{equation*}%
By assumption $d\{\cdot ,\cdot \}$ is an induced metric, and so for vectors $%
x$ and $z$ ({of the }same dimension) $d\{x,z\}=0$ if and only if $x=z.$
Using this fact we see that 
\begin{equation*}
d\{\mathbf{b}(\mathbf{\theta }^{0}),\mathbf{b}(\mathbf{\theta })\}=0\iff
\left( 
\begin{array}{c}
\mathbf{b}_{1}\mathbf{(\theta }^{0}\mathbf{)} \\ 
\mathbf{b}_{2}\mathbf{(\theta }^{0}\mathbf{)}%
\end{array}%
\right) =\left( 
\begin{array}{c}
\mathbf{b}_{1}\mathbf{(\theta )} \\ 
\mathbf{b}_{2}\mathbf{(\theta )}%
\end{array}%
\right) .
\end{equation*}%
The key observation is that the set $S_{\mathbf{\theta }}:=\left\{ \theta
\in \Theta :\Vert \mathbf{b}_{2}(\mathbf{\theta }^{0})-\mathbf{b}_{2}(%
\mathbf{\theta })\Vert =0\right\} $ always includes the point $\mathbf{%
\theta =\theta }^{0}$, but can include other points since $\mathbf{b}_{2}%
\mathbf{(\cdot )}$ need not be one-to-one. However, by [C2(2)], we know that 
$\mathbf{b}_{1}(\mathbf{\cdot })$ is one-to-one and so the only value of $%
\mathbf{\theta }$ for which $\qquad \qquad \qquad \qquad \qquad \qquad
\qquad \qquad $%
\begin{equation*}
\left( 
\begin{array}{c}
\mathbf{b}_{1}\mathbf{(\theta }^{0}\mathbf{)} \\ 
\mathbf{b}_{2}\mathbf{(\theta }^{0}\mathbf{)}%
\end{array}%
\right) -\left( 
\begin{array}{c}
\mathbf{b}_{1}\mathbf{(\theta )} \\ 
\mathbf{b}_{2}\mathbf{(\theta )}%
\end{array}%
\right) =\left( 
\begin{array}{c}
0 \\ 
0%
\end{array}%
\right) ,
\end{equation*}%
is $\mathbf{\theta =\theta }^{0}.$ The result then follows by the same
arguments as in Theorem \ref{thm1}.
\end{proof}

\bigskip

\begin{proof}[Proof of Theorem 2]
By the triangle inequality 
\begin{equation}
d\{\mathbf{\widehat{\beta }}(\mathbf{y}),\mathbf{\widehat{\beta }}(\mathbf{z}%
^{i})\}\leq d\{\mathbf{\widehat{\beta }}(\mathbf{y}),\mathbf{b}(\mathbf{%
\theta }^{i})\}+d\{\mathbf{b}(\mathbf{\theta }^{i}),\mathbf{\widehat{\beta }}%
(\mathbf{z}^{i})\}.  \label{beta1}
\end{equation}%
Before proceeding further we must show that, under the maintained
assumptions,\footnote{%
Recall that $d\{\cdot ,\cdot \}$ is an induced metric and hence convergence
in $\Vert \cdot \Vert $ will imply convergence in $d\{\cdot ,\cdot \}$.} 
\begin{equation}
\sup_{\mathbf{\theta \in \Theta }}\left\Vert \mathbf{\widehat{\beta }}(%
\mathbf{z}(\mathbf{\theta }))-\mathbf{b}(\mathbf{\theta })\right\Vert
=o_{P}(1).  \label{z_op1}
\end{equation}%
Define the following terms: 
\begin{eqnarray*}
\widetilde{Q}(\mathbf{\theta ,\beta }) &=&Q(\mathbf{z}(\mathbf{\theta }),%
\mathbf{\beta })-Q_{\infty }(\mathbf{\theta },\mathbf{b}(\mathbf{\theta })),
\\
\widetilde{Q}_{\infty }(\mathbf{\theta ,\beta }) &=&Q_{\infty }(\mathbf{%
\theta ,\beta })-Q_{\infty }(\mathbf{\theta },\mathbf{b}(\mathbf{\theta })).
\end{eqnarray*}%
Note that, by [G1(1)], for all $\delta >0$, if $\sup_{\mathbf{\theta \in
\Theta }}\Vert \widehat{\mathbf{\beta }}(\mathbf{z}(\mathbf{\theta )})-%
\mathbf{b}(\mathbf{\theta })\Vert >\delta ,$ there exists $\epsilon (\delta
)>0$, such that 
\begin{equation*}
\sup_{\mathbf{\theta \in \Theta }}\Vert Q_{\infty }(\mathbf{\theta },%
\widehat{\mathbf{\beta }}(\mathbf{z}(\mathbf{\theta )}))-Q_{\infty }(\mathbf{%
\theta },\mathbf{b}(\mathbf{\theta }))\Vert =\sup_{\mathbf{\theta \in \Theta 
}}\Vert \widetilde{Q}_{\infty }(\mathbf{\theta },\widehat{\mathbf{\beta }}(%
\mathbf{z}(\mathbf{\theta )}))\Vert >\epsilon (\delta ).
\end{equation*}%
From here, note that 
\begin{equation*}
\text{Pr}\left( \sup_{\mathbf{\theta \in \Theta }}\Vert \widehat{\mathbf{%
\beta }}(\mathbf{z}(\mathbf{\theta )})-\mathbf{b}(\mathbf{\theta })\Vert
>\delta \right) \leq \text{Pr}\left( \sup_{\mathbf{\theta \in \Theta }}\Vert 
\widetilde{Q}_{\infty }(\mathbf{\theta },\widehat{\mathbf{\beta }}(\mathbf{z}%
(\mathbf{\theta )}))\Vert >\epsilon (\delta )\right) .
\end{equation*}%
The result in (\ref{z_op1}) then follows if $\sup_{\mathbf{\theta \in \Theta 
}}\Vert \widetilde{Q}_{\infty }(\mathbf{\theta },\widehat{\mathbf{\beta }}(%
\mathbf{z}(\mathbf{\theta )}))\Vert =o_{P}(1)$.

Uniformly in $\mathbf{\theta }$, 
\begin{eqnarray}
\Vert \widetilde{Q}_{\infty }(\mathbf{\theta },\widehat{\mathbf{\beta }}(%
\mathbf{z}(\mathbf{\theta )})))\Vert &\leq &\Vert \widetilde{Q}_{\infty }(%
\mathbf{\theta },\widehat{\mathbf{\beta }}(\mathbf{z}(\mathbf{\theta )}))-%
\widetilde{Q}_{{}}(\mathbf{\theta },\widehat{\mathbf{\beta }}(\mathbf{z}(%
\mathbf{\theta )}))\Vert +\Vert \widetilde{Q}_{{}}(\mathbf{\theta },\widehat{%
\mathbf{\beta }}(\mathbf{z}(\mathbf{\theta )}))\Vert  \notag \\
&=&\Vert Q_{\infty }(\mathbf{\theta ,}\widehat{\mathbf{\beta }}(\mathbf{z}(%
\mathbf{\theta )}))-Q(\mathbf{z}(\mathbf{\theta ),}\widehat{\mathbf{\beta }}(%
\mathbf{z}(\mathbf{\theta )}))\Vert +\Vert \widetilde{Q}_{{}}(\mathbf{\theta 
},\widehat{\mathbf{\beta }}(\mathbf{z}(\mathbf{\theta )}))\Vert  \notag \\
&\leq &\sup_{\mathbf{\beta }\in \mathbf{B}}\Vert Q_{\infty }(\mathbf{\theta }%
,\mathbf{\beta })-Q(\mathbf{z}(\mathbf{\theta }),\mathbf{\beta })\Vert
+\Vert \widetilde{Q}(\mathbf{\theta },\widehat{\mathbf{\beta }}(\mathbf{z}(%
\mathbf{\theta )}))\Vert  \notag \\
&\leq &o_{P}(1)+\Vert \widetilde{Q}(\mathbf{\theta },\widehat{\mathbf{\beta }%
}(\mathbf{z}(\mathbf{\theta )}))\Vert .  \label{supb1}
\end{eqnarray}%
The first inequality follows from the {triangle inequality, the second from
the }definition of $\widetilde{Q}_{\infty }(\mathbf{\theta },\mathbf{\beta }%
) $ and $\widetilde{Q}_{{}}(\mathbf{\theta },\mathbf{\beta })$, the {third}
from the definition of $\sup ,$ and the {last from} Assumption [G1(2)].

From \eqref{supb1}, the result follows if 
\begin{equation*}
\sup_{\mathbf{\theta }\in \mathbf{\Theta }}\Vert \widetilde{Q}(\mathbf{%
\theta },\widehat{\mathbf{\beta }}(\mathbf{z}(\mathbf{\theta })))\Vert
=o_{P}(1).
\end{equation*}%
By the definition of $\widehat{\mathbf{\beta }}(\mathbf{z}(\mathbf{\theta }%
)) $, uniformly in $\mathbf{\theta }$, \emph{\ } 
\begin{eqnarray}
\Vert \widetilde{Q}(\mathbf{\theta },\widehat{\mathbf{\beta }}(\mathbf{z}(%
\mathbf{\theta )}))\Vert &\leq &\inf_{\mathbf{\beta }\in \mathbf{B}}\Vert 
\widetilde{Q}(\mathbf{\theta },\mathbf{\beta })\Vert +o_{P}(1)  \notag \\
&\leq &\inf_{\mathbf{\beta }\in \mathbf{B}}\Vert \widetilde{Q}(\mathbf{%
\theta },\mathbf{\beta })-\widetilde{Q}_{\infty }(\mathbf{\theta },\mathbf{%
\beta })\Vert +\inf_{\mathbf{\beta }\in \mathbf{B}}\Vert \widetilde{Q}%
_{\infty }(\mathbf{\theta },\mathbf{\beta })\Vert +o_{P}(1)  \notag \\
&\leq &\sup_{\mathbf{\beta }\in \mathbf{B}}\Vert Q(\mathbf{z}(\mathbf{\theta 
}),\mathbf{\beta })-Q_{\infty }(\mathbf{\theta },\mathbf{\beta })\Vert
+0+o_{P}(1)  \notag \\
&\leq &o_{P}(1),  \label{supb2}
\end{eqnarray}%
{with the last inequality following from [G1(2)]. }Combining equations %
\eqref{supb1} and \eqref{supb2} yields $\sup_{\mathbf{\theta \in \Theta }%
}\left\Vert \mathbf{\widehat{\beta }}(\mathbf{z}(\mathbf{\theta }))-\mathbf{b%
}(\mathbf{\theta })\right\Vert =o_{P}(1),$ and we can conclude 
\begin{equation}
d\{\widehat{\mathbf{\beta }}(\mathbf{z}^{i}),\mathbf{b}(\mathbf{\theta }%
^{i})\}\leq \sup_{\mathbf{\theta }\in \mathbf{\Theta }}d\{\mathbf{\widehat{%
\beta }}(\mathbf{z}(\mathbf{\theta })),\mathbf{b}(\mathbf{\theta }%
)\}=o_{P}(1).  \label{unifbeta}
\end{equation}%
Applying equation \eqref{unifbeta} to equation \eqref{beta1} we have 
\begin{equation}
d\{\mathbf{\widehat{\beta }}(\mathbf{y}),\mathbf{\widehat{\beta }}(\mathbf{z}%
^{i})\}\leq d\{\mathbf{\widehat{\beta }}(\mathbf{y}),\mathbf{b}(\mathbf{%
\theta }^{i})\}+o_{P}(1).  \label{star_1}
\end{equation}%
Applying the triangle inequality to $d\{\mathbf{\widehat{\beta }}(\mathbf{y}%
),\mathbf{b}(\mathbf{\theta }^{i})\}$ yields 
\begin{equation*}
d\{\mathbf{\widehat{\beta }}(\mathbf{y}),\mathbf{b}(\mathbf{\theta }%
^{i})\}\leq d\{\mathbf{\widehat{\beta }}(\mathbf{y}),\mathbf{b}(\mathbf{%
\theta }^{0})\}+d\{\mathbf{b}(\mathbf{\theta }^{0}),\mathbf{b}(\mathbf{%
\theta }^{i})\}+o_{P}(1),
\end{equation*}%
By [G1] and [G2], $\left\Vert \mathbf{\widehat{\beta }}(\mathbf{y})-\mathbf{b%
}(\mathbf{\theta }^{0})\right\Vert =o_{P}(1),$ and so 
\begin{equation}
d\{\mathbf{\widehat{\beta }}(\mathbf{y}),\mathbf{b}(\mathbf{\theta }%
^{i})\}\leq d\{\mathbf{b}(\mathbf{\theta }^{0}),\mathbf{b}(\mathbf{\theta }%
^{i})\}+o_{P}(1).  \label{star_2}
\end{equation}%
From \eqref{star_1} and \eqref{star_2} we {thus }have 
\begin{equation*}
d\{\mathbf{\widehat{\beta }}(\mathbf{y}),\mathbf{\widehat{\beta }}(\mathbf{z}%
^{i})\}\leq d\{\mathbf{b}(\mathbf{\theta }^{0}),\mathbf{b}(\mathbf{\theta }%
^{i})\}+o_{P}(1).
\end{equation*}%
For fixed ${\varepsilon }\geq 0$, as $T\rightarrow \infty $ a value $\mathbf{%
\theta }^{i}$ will be selected if and only if 
\begin{equation*}
d\{\mathbf{b}(\mathbf{\theta }^{0}),\mathbf{b}(\mathbf{\theta }%
^{i})\}+o_{P}(1)\,\,\leq \varepsilon .
\end{equation*}%
By [G3] the only value of $\mathbf{\theta }^{i}\in \mathbf{\Theta }$ for
which\textbf{\ }$\mathbf{b}\left( \mathbf{\theta }^{i}\right) =\mathbf{b}%
\left( \mathbf{\theta }^{0}\right) $ is $\mathbf{\theta }^{i}\mathbf{=\theta 
}^{0}$. Therefore, the only value of $\mathbf{\theta }^{i}$ satisfying $d\{%
\mathbf{b}(\mathbf{\theta }^{0}),\mathbf{b}(\mathbf{\theta }^{i})\}\leq
\varepsilon $ as $\varepsilon \rightarrow 0$ is $\mathbf{\theta }^{i}\mathbf{%
=\theta }^{0}$. The result follows using similar arguments to those of
Theorem \ref{thm1}.
\end{proof}


\begin{thebibliography}{99}
\bibitem{} Beaumont, M.A., Cornuet, J-M., Marin, J-M. and Robert, C.P.%
\textit{\ }2009. Adaptive Approximate Bayesian Computation, \textit{%
Biometrika}, 96, 983--990.

\bibitem{} Beaumont, M.A., Zhang, W. and Balding, D.J. 2002. Approximate
Bayesian Computation in Population Genetics, \textit{Genetics}, 162,
2025--2035.

\bibitem{} Beck, J.V. and Arnold, K.J. 1977. Parameter Estimation in
Engineering and Science, \textit{John Wiley \& Sons}.

\bibitem{} Biau, G., C\'{e}rou, F. and Guyader, A. 2015. New insights into
Approximate Bayesian Computation. \textit{Ann. Inst. H. Poincar\'{e} Probab.
Statist.}, 51, 376--403.

\bibitem{} Blackwell, D. and Dubins, L. 1962. Merging of Opinions with
Increasing Information, \textit{Ann. Statist.}, 33, 882--886.

\bibitem{} Blum, M.G.B. 2010. Approximate Bayesian Computation: a
Nonparametric Perspective, \textit{Journal of the American Statistical
Association}, 105, 1178-1187.

\bibitem{} Blum, M.G.B. and Fran\c{c}ois, O. 2010. Non-linear Regression
Models for Approximate Bayesian Computation, \textit{Statistics and Computing%
}, 20, 63--73.

\bibitem{} Blum, M.G.B., Nunes, M.A., Prangle, D. and Sisson, S.A. 2013. A
Comparative Review of Dimension Reduction Methods in Approximate Bayesian
Computation, \textit{Statistical Science}, 28, 189--208.

\bibitem{} Creel, M. and Kristensen, D. 2015. ABC of SV: Limited Information
Likelihood Inference in Stochastic Volatility Jump-Diffusion Models, \textit{%
Journal of Empirical Finance,} 31, 85-108.

\bibitem{} Diaconis, P. and Freedman, D. 1986. On the Consistency of Bayes
Estimates, \textit{Ann. Statist.}, Vol 14, no. 1, 1--26.

\bibitem{} Drovandi, C.C., Pettitt, A.N. and Faddy, M.J. 2011. Approximate
Bayesian Computation Using Indirect Inference, \textit{J. Royal Statistical
Soc. Series C}, 60 1 --21.

\bibitem{} Drovandi, C. C., Pettitt, A. N. and Lee, A. 2015. Bayesian
Indirect Inference using a Parametric Auxiliary Model. \textit{Statistical
Science}, 30, 72--95.

\bibitem{} Duffie, D. and Singleton, K.J. 1993. Simulated Moments Estimation
of Markov Models of Asset Prices, \textit{Econometrica}, 64, 929--952.

\bibitem{} Fearnhead, P. and Prangle, D. 2012. Constructing Summary
Statistics for Approximate Bayesian Computation: Semi-automatic Approximate
Bayesian Computation. \textit{J. Royal Statistical Soc. Series B}, 74,
419--474.

\bibitem{} Gallant, A.R. and Tauchen, G. 1996. Which Moments to Match, 
\textit{Econometric Theory}, 12, 657--681.

\bibitem{} Ghosal, S., Ghosh, J.K. and Samanta, T. 1995. On Convergence of
Posterior Distributions. \textit{Ann. Statist.}, 23, 2145--2152.

\bibitem{} Gleim, A. and Pigorsch, C. 2013. Approximate Bayesian Computation
with Indirect Summary Statistics. \textit{Draft paper:} \textit{%
http://ect-pigorsch.mee.uni-bonn.de/data/research/papers/}.

\bibitem{} Gouri\'{e}roux, C. and Monfort, A. 1996. \textit{Simulation-based
Econometric Methods}, OUP.

\bibitem{} Gouri\'{e}roux, C., Monfort, A. and Renault, E. 1993. Indirect
Inference, \textit{Journal of Applied Econometrics}, 85, S85--S118.

\bibitem{} Heggland, K. and Frigessi, A. 2004. Estimating Functions in
Indirect Inference, \textit{J. Royal Statistical Soc. Series B}, 66,
447--462.

\bibitem{} Ibragimov, I. A. and Has'minskii, R. Z. 1981. \textit{Statistical
Estimation: Asymptotic Theory}. Springer, New York.

\bibitem{} Jennrich, R. I. 1969. Asymptotic Properties of Non-Linear Least
Squares Estimators. \textit{Ann. Math. Statist.}, 40, 633--643.

\bibitem{} Joyce, P. and Marjoram, P. 2008. Approximately Sufficient
Statistics and Bayesian Computation. \textit{Statistical applications in
genetics and molecular biology}, 7, 1--16.

\bibitem{} Le Cam. 1953. On Some Asymptotic Properties of Maximum Likelihood
Estimates and Related Bayes Estimates. \textit{University of California
Publications in Statistics}, 1, 277--330.

\bibitem{} Li, W. and Fearnhead, P. 2015. Behaviour of ABC for Big Data, 
\textit{http://arxiv.org/abs/1506.03481.}

\bibitem{} Marin, J-M., Pudlo, P., Robert, C.P. and Ryder, R. 2011.
Approximate Bayesian Computation Methods. \textit{Statistics and Computing},
21, 289--291.

\bibitem{} Marin, J-M., Pillai, N., Robert, C.P. and Rousseau, J. 2014.
Relevant statistics for Bayesian model choice. \textit{J. Royal Statistical
Soc. Series B}, 76, 833--859.

\bibitem{} Martin, G.M., McCabe, B.P.M., Maneesoonthorn, O. and Robert, C.P.
2014. Approximate Bayesian Inference in State Space Models, \textit{%
http://arxiv.org/abs/1409.8363.}

\bibitem{} Martin, J.S., Jasra, A., Singh, S.S., Whiteley, N., Del Morale,
P. and McCoy, E. 2014. Approximate Bayesian Computation for Smoothing. 
\textit{Stochastic Analysis and Applications, }32, 397-420.

\bibitem{} Marjoram, P., Molitor, J., Plagonal, V. and Tavar\'{e}, S. 2003.
Markov Chain Monte Carlo Without Likelihoods, \textit{Proceedings of the
National Academie of Science USA}, 100, 15324--15328.

\bibitem{} McFadden, D. 1989. A Method of Simulated Moments for Estimation
of Discrete Response Models Without Numerical Integration, \textit{%
Econometrica}, 57, 995--1026.

\bibitem{} Milstein, G. 1978. A Method of Second Order Accuracy Integration
of Stochastic Differential Equations, \textit{Theory of Probability and Its
Applications}, 23, 396--401.

\bibitem{} Newey, W.K. and McFadden, D. 1994. Large Sample Estimation and
Hypothesis Testing, In \textit{Handbook of Econometrics }(Eds. Engle and
McFadden), Amsterdam: Elsevier Science.

\bibitem{} Nott D., Fan, Y., Marshall, L. and Sisson, S. 2014. Approximate
Bayesian Computation and Bayes Linear Analysis: Towards High-dimensional
ABC, \textit{Journal of Computational and Graphical Statistics}, 23, 65--86.

\bibitem{} Pakes, A. and Pollard, D. 1989. Simulation and the Asymptotics of
Optimization Estimators, \textit{Econometrica}, 57, 1027--1057.

\bibitem{} Pollard , D. 1990. Empirical Processes: Theory and Applications.
NSF-CBMS Regional Conference Series in Probability and Statistics, 2, 1--86.

\bibitem{} Prangle, D. 2015. Adapting the ABC distance function,\textit{\
http://arxiv.org/pdf/1507.00874.}

\bibitem{} Pritchard, J.K., Seilstad, M.T., Perez-Lezaun, A. and Feldman,
M.W. 1999. Population Growth of Human Y Chromosomes: A Study of Y Chromosome
Microsatellites, \textit{Molecular Biology and Evolution}, 16, 1791--1798.

\bibitem{} Sisson S. and Fan, Y. 2011. Likelihood-free Markov Chain Monte
Carlo. In \textit{Handbook of Markov Chain Monte Carlo} (Eds. Brooks,
Gelman, Jones, Meng). Chapman and Hall/CRC Press.

\bibitem{} Sisson, S., Fan, Y. and Tanaka, M. 2007. Sequential Monte Carlo
without Likelihoods, \textit{Proceedings of the National Academie of Science
USA}, 104, 1760--1765.

\bibitem{} Sun, L., Lee, C. and Hoeting, J.A. 2014. Parameter Inference and
Model Selection in Deterministic and Stochastic Dynamical Models via
Approximate Bayesian Computation: Modeling a Wildlife Epidemic,
http://arxiv.org/pdf/1409.7715.pdf.

\bibitem{} Tavar\'{e}, S., Balding, D.J., Griffiths, R.C. and Donnelly, P.
1997. Inferring Coalescence Times from DNA Sequence Data, \textit{Genetics},
145, 505--518.

\bibitem{} Toni, T., Welch, D., Strelkowa, N., Ipsen, A. and Stumpf, M.P.H.
2009. Approximate Bayesian Computation Scheme for Parameter Inference and
Model Selection in Dynamical Systems, \textit{JRSS (Interface)}, 6, 187--202.

\bibitem{} Wegmann, D., Leuenberger, C. and Excoffier, L. 2009. Efficient
Approximate Bayesian Computation Coupled with Markov chain Monte Carlo with
Likelihood, \textit{Genetics}, 182, 1207--1218.
\end{thebibliography}
\end{document}


\title{\textbf{On Consistency of Approximate Bayesian Computation:
Supplementary Appendix }}
\author{David T. Frazier, Gael M. Martin and Christian P. Robert}
\maketitle

\begin{abstract}
The definition of {Bayesian consistency} used in {the} paper holds when $%
T\rightarrow \infty $ and $\varepsilon \rightarrow 0$, with no specific rate
condition between $T$ and $\varepsilon $ being required. To understand why,
in general contexts, the way in which $T$ and $\varepsilon $ move toward
their limits does not matter readers are referred to the proof of Theorem 1.
However, for illustrative purposes, we consider {here }a simple analytic
example {that }demonstrat{es} this point {explicitly}. The example is
similar in spirit, but not substance, to examples considered in both
Wilkinson (2013) and Li and Fearnhead (2015). {The contrast between the
situation that obtains in proving Bayesian consistency and the need for a }$%
T $-{dependent condition for }$\varepsilon $ {when proving the asymptotic
normality of ABC point estimators, is also highlighted.}
\end{abstract}

\section{Bayes Consistency Demonstration}

\noindent Bayesian consistency of ABC\ is determined by the behavior of
objects of the form%
\begin{equation*}
\text{Pr}_{\varepsilon }(\mathbf{\theta }\in \Lambda _{\delta }|\mathbf{\eta
(y)})=\int_{\Lambda _{\delta }}p_{\varepsilon }(\mathbf{\theta }|\mathbf{%
\eta (y))}d\mathbf{\theta },
\end{equation*}%
where $\Lambda _{\delta }=\Theta /\mathcal{\aleph }_{\delta }(\mathbf{\theta 
}^{0})$ and $\mathcal{\aleph }_{\delta }(\mathbf{\theta }^{0})$ is a delta
neighborhood of $\mathbf{\theta }^{0}$, for any $\delta >0$. In particular,
the property we prove is 
\begin{equation*}
\lim_{T\rightarrow \infty }\lim_{\varepsilon \rightarrow 0}\text{Pr}\left( 
\text{Pr}_{\varepsilon }(\mathbf{\theta }\in \Lambda _{\delta }|\mathbf{\eta
(y)})>\xi \right) =0
\end{equation*}%
for any $\xi >0$ and arbitrarily small. Therefore, if for any\textbf{\ }$%
\delta >0$, Pr$_{\varepsilon }(\mathbf{\theta }\in \Lambda _{\delta }|%
\mathbf{\eta (y)})=o_{P}(1),$ as $T\rightarrow \infty $ and $\varepsilon
\rightarrow 0$, the result follows.{\ As the proof of Theorem 1
demonstrates, the conditions [S0]-[S2(2)] ensure that this is the case, with
no rate on }$\varepsilon $ {that is }$T$-{dependent required.}

To {further }cultivate intuition {as to }why a specific rate condition
between $T$ and $\varepsilon $ is not required, let us consider the
following simple example. Assume we are interested in inference on the {%
scalar }mean parameter in a Gaussian {model, }$y_{i}\sim _{i.i.d.}N(\theta
^{0},1)$, using ABC, {with }observed data $\mathbf{y}%
=(y_{1},y_{2},...,y_{T})^{\prime }$. To this end, we take as our summary
statistic the sample mean $\eta (\mathbf{y})=\frac{1}{T}\sum_{t=1}^{T}y_{t}.$
Letting $\eta (\mathbf{z(}\theta \mathbf{)})$ denote the simulated version
of $\eta (\mathbf{y})$, we follow Wilkinson (2013) and assume {that }for
some value $\bar{\theta}$ the observed summary statistic satisfies 
\begin{equation*}
\eta (\mathbf{y})=\eta (\mathbf{z(}\bar{\theta}\mathbf{)})+v,\;\;v\sim
N(0,\varepsilon ^{2}).
\end{equation*}%
We then accept draws $\theta ^{(i)}$ for use in building $%
p_{\varepsilon}(\theta |\eta (\mathbf{y}))$ according to {Algorithm \ref%
{ar1_ABC}.}

\begin{algorithm}
\caption{ABC algorithm}\label{ar1_ABC}
\begin{algorithmic}[1]
\State Simulate $\theta ^{i}$, $i=1,2,...,R$, from $p(\theta )\sim
N(0,1)$;
\State Simulate $\eta (\mathbf{z}^{i})$, $i=1,2,...,R$, from the
likelihood, $p(\mathbf{.|}\theta ^{i})$;
\State Accept $\theta ^{i}$ with probability:%
\begin{equation}
K_{\varepsilon }\left( \eta (\mathbf{y})-\eta (\mathbf{z}(\theta
^{(i)}))\right) =\frac{1}{\sqrt{2\pi \varepsilon ^{2}}}\exp \left( -\frac{%
\left( \eta (\mathbf{y})-\eta (\mathbf{z}(\theta ^{(i)}))\right) ^{2}}{%
\varepsilon ^{2}}\right) ,  \label{distance}
\end{equation}%
where $\varepsilon $ plays the role of the bandwidth in the Gaussian kernel.
\end{algorithmic}
\end{algorithm}

\bigskip Noting that $\eta (\mathbf{z}(\theta ))\sim N(\theta ,\frac{1}{T}),$
by Theorem 1 of Wilkinson (2013) we have that 
\begin{equation*}
p_{\varepsilon }(\theta |\eta (\mathbf{y}))\propto N\left( \frac{\eta (%
\mathbf{y})}{(1/T)+\varepsilon ^{2}+1},\frac{1+T\varepsilon ^{2}}{%
T+1+T\varepsilon ^{2}}\right) .
\end{equation*}%
As highlighted above, Bayesian consistency is only concerned with the limit
of the following probability 
\begin{equation}
\text{Pr}_{\epsilon }(\theta \in \Lambda _{\delta }|\eta (\mathbf{y}%
))=\int_{\Lambda _{\delta }}p_{\varepsilon }(\theta |\eta \mathbf{(y))}%
d\theta \equiv \int_{-\infty }^{\theta ^{0}-\delta }p_{\varepsilon }(\theta
|\eta (\mathbf{y}))d\theta +\int_{\theta ^{0}+\delta }^{\infty
}p_{\varepsilon }(\theta |\eta (\mathbf{y}))d\theta .  \label{prob}
\end{equation}%
For this example, the probability {in (\ref{prob})} can be calculated
analytically {as} 
\begin{eqnarray}
\text{Pr}_{\epsilon }(\theta \in \Lambda _{\delta }|\eta (\mathbf{y})) &=&-%
\frac{\sqrt{2}\,\left( \mathop{\mathrm{Erf}}\nolimits\!\left( \sqrt{\frac{T\,%
{\varepsilon }^{2}+T+1}{T\,{\varepsilon }^{2}+1}}\,\left( \delta -\theta
^{0}+\frac{\eta (\mathbf{y})}{{\varepsilon }^{2}+\frac{1}{T}+1}\right)
\right) -1\right) }{4}  \notag \\
&&-\frac{\sqrt{2}\,\left( \mathop{\mathrm{Erf}}\nolimits\!\left( \sqrt{\frac{%
T\,{\varepsilon }^{2}+T+1}{T\,{\varepsilon }^{2}+1}}\,\left( \delta +\theta
^{0}-\frac{{\eta (\mathbf{y})}}{{\varepsilon }^{2}+\frac{1}{T}+1}\right)
\right) -1\right) }{4}  \label{prob2}
\end{eqnarray}%
where 
\begin{equation*}
\text{Erf}\left( x\right) =\frac{2}{\sqrt{\pi }}\int_{0}^{x}\exp \left(
-t^{2}\right) dt.
\end{equation*}%
Note that, from its definition, $-1\leq \text{Erf}(x)\leq 1$ for all $x\in 
\mathbb{R}$. For simplicity, let us re-define the terms on the RHS of %
\eqref{prob2} inside $\text{Erf}(\cdot )$ as%
\begin{eqnarray*}
x_{1}(T,\varepsilon ) &=&\left( \sqrt{\frac{T\,{\varepsilon }^{2}+T+1}{T\,{%
\varepsilon }^{2}+1}}\,\left( \delta -\theta ^{0}+\frac{\eta (\mathbf{y})}{{%
\varepsilon }^{2}+\frac{1}{T}+1}\right) \right) \text{ and} \\
x_{2}(T,\varepsilon ) &=&\left( \sqrt{\frac{T\,{\varepsilon }^{2}+T+1}{T\,{%
\varepsilon }^{2}+1}}\,\left( \delta +\theta ^{0}-\frac{\eta (\mathbf{y})}{{%
\varepsilon }^{2}+\frac{1}{T}+1}\right) \right) .
\end{eqnarray*}%
{Given that }$\lim_{x\rightarrow \infty }$Erf$(x)=1$, {it follows that }$%
\text{Pr}_{\epsilon }(\theta \in \Lambda _{\delta }|\eta (\mathbf{y}))%
\overset{P}{\rightarrow }0$ if%
\begin{equation*}
\lim_{T\rightarrow \infty }\lim_{\varepsilon \rightarrow
0}x_{j}(T,\varepsilon )\overset{P}{\rightarrow }\infty \text{; }j=1,2.
\end{equation*}

\vspace{.25cm}

\noindent \textbf{Lemma:} $\text{Pr}_{\epsilon }(\theta \in \Lambda _{\delta
}|\eta (\mathbf{y}))\overset{P}{\rightarrow }0$ as $T\rightarrow \infty $
and $\varepsilon \rightarrow 0$, with no particular rate condition between $%
T $ and $\varepsilon $ being required.

\vspace{.25cm}

The above {lemma} states that a sequential asymptotic analysis suffices for
Bayesian consistency. By sequential asymptotics, we mean first taking limits
with respect to $T\rightarrow \infty $ and then $\varepsilon \rightarrow 0$,
or $\varepsilon \rightarrow 0$ and then $T\rightarrow \infty $. In this
simple example we {prove this lemma} in two steps: first, by showing that $%
\text{Pr}_{\epsilon }(\theta \in \Lambda _{\delta }|\eta (\mathbf{y}))%
\overset{P}{\rightarrow }0$ when we let $\varepsilon \rightarrow 0$ and then 
$T\rightarrow \infty $; second, by considering the reverse scenario of $%
T\rightarrow \infty $ and then $\varepsilon \rightarrow 0$.

\vspace{.5cm}

\noindent (1) \textbf{Limits:} $\varepsilon \rightarrow 0$ and then $%
T\rightarrow \infty \bigskip $

\noindent First, let us consider taking the limit of the {first term on the
RHS of (\ref{prob2})} with respect to $\varepsilon \rightarrow 0$, {for
fixed }$T$. This limit is clearly determined by 
\begin{eqnarray*}
\lim_{\varepsilon \rightarrow 0}x_{1}(T,\varepsilon ) &=&\lim_{\varepsilon
\rightarrow 0}\left( \sqrt{\frac{T\,{\varepsilon }^{2}+T+1}{T\,{\varepsilon }%
^{2}+1}}\,\left( \delta -\theta ^{0}+\frac{\eta (\mathbf{y})}{{\varepsilon }%
^{2}+\frac{1}{T}+1}\right) \right) \\
&=&\left( \sqrt{\frac{T+1}{1}}\,\left( \delta -\theta ^{0}+\frac{\eta (%
\mathbf{y})}{\frac{1}{T}+1}\right) \right) .
\end{eqnarray*}%
Now, taking limits of $x_{1}(T,0)$ with respect to $T$ yields 
\begin{eqnarray*}
\plim_{T\rightarrow \infty }\left( \sqrt{\frac{T+1}{1}}\,\left( \delta
-\theta ^{0}+\frac{\eta (\mathbf{y})}{\frac{1}{T}+1}\right) \right) &=&\plim%
_{T\rightarrow \infty }\left\{ \left( \delta -\theta ^{0}+\frac{\eta (%
\mathbf{y})}{\frac{1}{T}+1}\right) \right\} \plim_{T\rightarrow \infty
}\left\{ \sqrt{T+1}\right\} \\
&=&\delta \lim_{T\rightarrow \infty }\sqrt{T+1}\rightarrow \infty ,
\end{eqnarray*}%
where we have used the fact that $\text{plim}_{T\rightarrow \infty }\eta (%
\mathbf{y})=\theta ^{0}$. We can therefore conclude that the first term on
the RHS of \eqref{prob2} is $o_{P}(1)$. {Taking the same steps, we find}%
\begin{equation*}
\lim_{T\rightarrow \infty }\lim_{\varepsilon \rightarrow
0}x_{2}(T,\varepsilon )\overset{P}{\rightarrow }\infty
\end{equation*}%
also, in which case the second term on the RHS of (\ref{prob2}) is $o_{P}(1)$%
. {No condition between }$T${\ and }$\varepsilon ${\ is used to obtain this
result.}

\vspace{.5cm}

\noindent (2) \textbf{Limits:} $T\rightarrow \infty $ and then $\varepsilon
\rightarrow 0\bigskip $

\noindent Now, let us consider the reverse situation where $T\rightarrow
\infty $ and then $\varepsilon \rightarrow 0$.\emph{\ }For fixed $%
\varepsilon \geq 0$, 
\begin{eqnarray*}
\plim_{T\rightarrow \infty }x_{1}(T,\varepsilon ) &=&\plim_{T\rightarrow
\infty }\,\left( \delta -\theta ^{0}+\frac{\eta (\mathbf{y})}{{\varepsilon }%
^{2}+\frac{1}{T}+1}\right) \lim_{T\rightarrow \infty }\sqrt{\frac{T\,{%
\varepsilon }^{2}+T+1}{T\,{\varepsilon }^{2}+1}} \\
&=&\left( \delta -\theta ^{0}+\frac{\theta ^{0}}{{\varepsilon }^{2}+1}%
\right) \sqrt{\frac{\varepsilon ^{2}+2}{\varepsilon ^{2}}},
\end{eqnarray*}%
{where }{L}'h\^{o}pital's rule {is used in the production of the second
equality.} From here we have that 
\begin{eqnarray*}
\lim_{\varepsilon \rightarrow 0}\left\{ \plim_{T\rightarrow \infty }\left( 
\sqrt{\frac{T\,{\varepsilon }^{2}+T+1}{T\,{\varepsilon }^{2}+1}}\,\left(
\delta -\theta ^{0}+\frac{\eta (\mathbf{y})}{{\varepsilon }^{2}+\frac{1}{T}+1%
}\right) \right) \right\} &=&\left\{ \delta -\theta ^{0}+\lim_{\varepsilon
\rightarrow 0}\frac{\theta ^{0}}{\varepsilon ^{2}+1}\right\} \left\{
\lim_{\varepsilon \rightarrow 0}\sqrt{\frac{\varepsilon ^{2}+2}{\varepsilon
^{2}}}\right\} \\
&=&\delta \left\{ \lim_{\varepsilon \rightarrow 0}\frac{\varepsilon ^{2}+2}{%
\varepsilon ^{2}}\right\} ^{1/2}\rightarrow \infty
\end{eqnarray*}%
{By similar arguments it follows that}%
\begin{equation*}
\lim_{\varepsilon \rightarrow 0}\left\{ \plim_{T\rightarrow \infty }\left( 
\sqrt{\frac{T\,{\varepsilon }^{2}+T+1}{T\,{\varepsilon }^{2}+1}}\,\left(
\delta +\theta ^{0}-\frac{\eta (\mathbf{y})}{{\varepsilon }^{2}+\frac{1}{T}+1%
}\right) \right) \right\} =\delta \left\{ \lim_{\varepsilon \rightarrow 0}%
\frac{\varepsilon ^{2}+2}{\varepsilon ^{2}}\right\} ^{1/2}\rightarrow \infty
\end{equation*}%
{and it follows once again that} 
\begin{equation*}
\lim_{\varepsilon \rightarrow 0}\left\{ {\plim}_{T\rightarrow \infty }\text{%
Pr}_{\varepsilon }(\theta \in \mathbf{\Lambda }_{\delta }|\eta (\mathbf{y}%
))\right\} =0.
\end{equation*}%
Again, no condition between $T$ and $\varepsilon $ {is invoked in producing}
this result.\bigskip

\noindent \textbf{Remark 1:} This simple example illustrates that the
concept of Bayesian consistency does not necessarily require a specific rate
condition between $T$ and $\varepsilon $, {with} sequential asymptotic
analysis {sufficing}. The Proof of Theorem 1 follows a very similar logic,
but requires more complicated arguments due to the generality of the objects
in question.\bigskip

\noindent \textbf{Remark 2:} It is also important to reiterate that while
our focus is on Bayesian consistency, other researchers have recently begun
to explore the asymptotic properties of ABC point {estimators}. For
example,\ Li and Fearnhead (2015) are concerned with asymptotic normality of
the ABC point {estimator}, or some function thereof, {and the conditions
invoked therein are} {much} stronger than our conditions for Bayesian
consistency. {Specifically}, to ensure that the ABC point {estimator} is
asymptotically normal (indeed, asymptotically equivalent to the MLE of the
relevant parameter function, given the summary statistic - {denoted by }MLES 
{therein}), {Li and Fearnhead} require $\varepsilon =o(1/\sqrt{T})$. This {%
order condition} {ensures} that the leading remainder term in the
first-order Taylor series expansion used by {the authors} to equate the ABC
posterior mean with the MLES is{\ $o_{P}(1/\sqrt{T}).${\ To see this, we
note that }from the proof of Theorem 3.1 in Li and Fearnhead {(see p. 23)}, }%
${\varepsilon }${\ {appears} as a square, which ensures that, if }$%
\varepsilon =$ {$o(1/\sqrt{T})$, the leading remainder term in the Taylor
expansion is 
\begin{equation*}
o_{P}(1/\sqrt{T})+r_{1}(\eta (y),T)\varepsilon ^{2}=o_{P}(1/\sqrt{T}%
)+r_{1}(\eta (y),T)o(1/T)\equiv o_{P}(1/\sqrt{T}),
\end{equation*}%
}as stated. Given that the goal of {our paper} is Bayesian consistency of
ABC, there is no reason to believe, \textit{a priori}, that the two sets of
assumptions should coincide.